\documentclass[11pt]{article}
\usepackage[utf8]{inputenc}
\usepackage[T1]{fontenc}
\usepackage{lmodern}
\usepackage{textcomp}
\usepackage{microtype}
\usepackage{geometry}
\usepackage{hyperref}
\hypersetup{hidelinks}
\usepackage{graphicx}
\usepackage{booktabs}
\usepackage{multirow}
\usepackage{tabularx}
\usepackage{array}
\usepackage{xcolor}
\usepackage{enumitem}
\usepackage{amsmath,amssymb}
\usepackage{caption}
\usepackage{subcaption}
\usepackage{tikz}
\usepackage{float} 
\usetikzlibrary{arrows.meta,positioning,fit,shapes.misc}
\usepackage[numbers]{natbib}

\title{An Explainable Agentic System for Detection of Conversational Scams with Summary-Based Memory}

\author{
Ahmed Omar Salim Adnan \\
University of New Haven \\
M.S. in Data Science \\
\texttt{aadna1@unh.newhaven.edu}
\and
Yogananda Manjunath \\
University of New Haven \\
M.S. in Data Science \\
\texttt{ymanj2@unh.newhaven.edu}
\and
Shivanjali Khare \\
University of New Haven \\
Department of Electrical \& \\Computer Engineering and Computer Science \\
\texttt{skhare@newhaven.edu}
}

\begin{document}
\maketitle

\begin{abstract}
	Following the rapid progress of generative Artificial Intelligence, there is a growing threat posed by conversational scams. These scams often span over multiple weeks or months, gradually build trust and request for money or sensitive information. Existing scam-detection systems mainly focus on isolated messages, which renders them inadequate against this evolving threat. This paper extends single-message phishing detection and presents an explainable agentic system for detecting sophisticated conversational scams. It also introduces ConScamBench-278, an initial public multi-category benchmark for conversational scam detection spanning eight scam types, released to support reproducible evaluation and future expansion. On isolated messages the single-message detector attains 100\% phishing recall, while the conversation-level detector identifies all conversational scams in the public LoveFraud02 corpus (83/83) and reaches 97.8\% accuracy (95\% CI [95.4, 99.0]) on ConScamBench-278. Two user studies ($N=100$ and $N=45$) further motivate the system: participants report frequently experiencing uncertainty when judging suspicious conversations. In an uncontrolled pre/post comparison, users  self-reported trust, self-confidence, and perceived need for AI-based scam detection all increased ($p<0.001$, Wilcoxon signed-rank). The system also receives a System Usability Scale score of 74.7 (95\% CI [72.5, 76.9]), above the established usability benchmark.
\end{abstract}

\section{Introduction}

The emergence of cyberscams is not sudden; rather, it evolved from traditional fraud schemes. With time, the digital aspect of scams proliferated in a variety of ways and became more convincing, as they rely on the same historic principles of psychological manipulation \cite{10.1093/bjc/azn074}. One of the most convincing cyberscams, phishing, refers to the fraudulent practice of sending messages, often in the false disguise of reputable companies. Here, the intention is to convince the victim to reveal sensitive personal information required to get access to their finances \cite{10.1145/1290958.1290968}. In 2024, phishing or spoofing was the most reported cybercrime to the FBI, totaling 193,407 complaints \cite{FBI2024InternetCrimeReport}.

In the modern era of computing and Artificial Intelligence (AI), these scams have evolved from simple email deception into multi-modal, AI-augmented attack ecosystems \cite{Hazell2023LLMPhishing}. Conversational scams are considered significantly more dangerous, as they can cause severe psychological trauma beyond financial harm \cite{Lea2009}. Hazell et al. led an extensive study on modern techniques of AI-based phishing attacks and found that AI-generated emails were often more persuasive than human-crafted ones. According to \cite{LetainMathieu2025AIGeneratedPhishing}, AI-generated phishing is the biggest enterprise threat of 2026, outpacing ransomware and insider risk. This concern is becoming more pressing as generative AI lowers the cost of producing fluent, personalized, and sustained scam conversations, allowing attackers to automate social-engineering messages that appear more natural than traditional template-based phishing. Given this progression, there is an immediate demand for equally strong countermeasures.

Existing AI-based phishing detectors typically focus on detecting phishing in single messages; we refer to such a tool as a Single-Message Detector (SMD). One such detection tool, SmishX, is a Large Language Model (LLM) based agentic phishing detector that takes a single message as input \cite{Wang2025SmishX}. Although quite effective, this system relies on multiple closed-source dependencies and cannot detect conversational scams. A further challenge is the availability of diverse datasets to evaluate conversation scam detection tools. Existing phishing benchmarks target malicious artifacts in short, single messages and do not capture the slower structure of conversational fraud. Public multi-turn scam datasets are much rarer, and when they do exist, they typically represent only one scam family.

This paper addresses these limitations by proposing an explainable agentic system for detecting scams across both isolated messages and multi-turn conversations. We use \emph{explainable} to denote architectural transparency together with evidence-based, user-facing rationale rather than formally faithful feature-level attribution. A rigorous evaluation of explanation quality is left to future work. The proposed system combines a SMD, sender-authenticity verification, an adaptive conversation-level investigator, summary-based memory, and a user-facing advice generator. Unlike detectors that classify each message independently, the proposed system maintains a compact representation of evolving conversations and uses it to identify delayed scam signals such as identity inconsistency, grooming behavior, redirection to external channels, urgency, secrecy, and financial requests.

The main contributions of this paper are as follows:
\begin{itemize}
	\item Proposes a modular explainable agentic architecture for conversational scam detection that integrates single-message artifact analysis with multi-turn conversation-level reasoning.
	\item Introduces a summary-based memory as a scalable alternative to repeatedly pass full conversation history into the detector, allowing long conversations to be analyzed with reduced context growth.
	\item Strengthens the sender-authenticity check by extracting and verifying URLs, phone numbers, email addresses, and claimed brand identities before final scam classification.
	\item Assembles and releases ConScamBench-278, one of the first multi-category benchmarks for multi-turn conversational scam detection. The dataset spans across eight scam types and combines real-world fraud with a minority of LLM-augmented adversarial simulations, reflecting emerging AI-assisted social engineering. The dataset enables evaluation across a far broader range of scam types than existing corpora allows.
	\item Evaluates the system comprehensively using single-message phishing data, the LoveFraud02 romance-fraud corpus, ConScamBench-278, a component ablation and a second-backend robustness check. Two user studies are performed to measure user uncertainty, trust, disagreement, and safety-oriented behavior, results are reported with confidence intervals and significance tests.
\end{itemize}

\section{Related Work}
\label{section_related}

Conversational scams can be of multiple types and are usually a long-term investment involving lengthy conversations rather than a single-message attack \cite{luo2026anatomy}. One type of conversational scam is ``Pig Butchering'' based on a translation of the Chinese term \textit{Sha Zhu Pan}, where the attacker fattens up the victim with emotional intimacy before butchering their savings. These attacks are carried out in three stages: hook, line, and sinker. The conversation usually starts with a wrong-number text or a professional-network request on an employment-oriented social networking service such as LinkedIn \cite{Acharya2024PigButcheringScams}.
Today, with easy access to AI tools, these attacks have become more intensive, and an attacker can maintain dozens of simultaneous, emotionally intelligent conversations. After long-term trust building, the scammer casually mentions lucrative crypto or gold trading, often referring to a fake trading dashboard \cite{Gressel2025RomanceBaitingLLM}. A common example of long-form of manipulation is the romance scam, where the scammer builds emotional closeness, invents a believable identity, and delays the financial request until the victim is invested in the relationship \cite{anesa2020lovextortion}. The final request, often framed as an emergency, can be travel trouble, customs fees, medical costs, a frozen account, or a sudden crisis involving a child or relative. Since the victim is already attached to the persona, such requests can seem credible even when they would sound implausible in isolation \cite{whitty2013scammers}. Furthermore, AI-era versions use generative AI to overcome language barriers and deepfake technology to bypass video-call verification. Unaware victims are very likely to fall for such high-quality AI-generated personas. In one notable case in 2025, an ``AI Brad Pitt'' scam duped a French woman out of \texteuro{}830,000 and demonstrated how AI can impersonate celebrities to manipulate victims over a long time period \cite{Tulga16022026}.

Work on dialogue systems shows that multi-turn understanding depends on memory: a record of what was said, by whom, which identity claims were made, and how the exchange shifts over time \cite{laban2026lost, maharana2024evaluating}. A system that treats every message as a fresh classification problem cannot see those signals. Scam detection inherits the problem. A request for money, secrecy, or contact on another channel is often suspicious because of where it falls in the conversation, not because of its wording \cite{whitty2013scammers}. A detector therefore needs memory that survives long chats. Feeding the entire transcript into every new decision, though, is expensive and unstable.

There is also significant concern regarding proprietary, closed-source LLMs in critical sectors such as cybersecurity. Issues regarding their unauthorized and prolonged retention of personal data and their lack of transparency throughout the pipeline are well documented \cite{tramer2025centralized}. SmishX \cite{Wang2025SmishX} depends heavily on third-party APIs, and in sensitive tasks such as scam detection, privacy matters more than in ordinary text classification. Conversations can contain personal histories, emotional disclosures, finances, family details, or sensitive account information.

LLMs also require a significant amount of high-quality varied data and rely heavily on data collection. The amount of data scraped from books, articles, and similar sources is not sufficient for further training and improving model predictions. As such, data is also collected from users during interactions in real time. This includes normal texts, user queries, and further information provided by the user in support of their question, up to personal records such as email addresses, phone numbers, and usage data \cite{Sebastian2023Privacy}. Additionally, general users of LLMs come from varied demographics, including vulnerable teens and seniors. Many do not have a proper understanding of what data are being collected and how they are handled \cite{Alzamil2025ChatGPTPrivacy}. Although organizations enforce strict data-storage practices, including encryption, deletion, and anonymization of data, there remains a substantial concern that these models memorize personal information during training. There are various ways that adversaries can access personal data memorized by these models. One technique, called ``training data extraction'' involves designing specific prompts and requesting the model to complete prompt sentences \cite{274574}, or prompting with context about personal-data patterns to retrieve memorized email addresses and other associated data as output \cite{huang-etal-2022-large}. Fine-tuning techniques can be used to fine-tune the full model, the model head, or an adapter in order to manipulate the model into leaking information that it otherwise would not \cite{mireshghallah-etal-2022-empirical}. Other methods, such as jailbreaking, can also assist in leaking information. An in-depth analysis of these attacks is out of scope for this paper.

Another concern is that the governing policies of proprietary LLM services are not stable from the perspective of system developers. The privacy policy, retention terms, acceptable-use restrictions, research-access conditions, and API usage rules are set by the provider and may be revised unilaterally. This creates a moving compliance target for any application built on top of such services. A recent analysis of the Terms of Service of major LLM providers notes both substantial variation across providers and a broader pattern of changing access conditions, including reduced researcher-specific access and shifts toward pay-to-play or restricted access models in what the authors describe as a ``post-API'' environment \cite{Davidson2026RegulatoryGrayAreas}. In other words, even if a system is compliant at deployment time, the surrounding contractual and governance conditions may later change in ways that affect how the system can be used, what data may be processed, or what assurances can reasonably be made to users. This instability is especially risky in sensitive applications such as scam detection, where the system may process sensitive information belonging to the user. If the service provider later changes its retention rules, training-related data handling, or permitted-use boundaries, the effective privacy and governance profile of the application may also change even though the application's architecture itself has not. Previous work on LLM privacy and governance emphasizes that privacy risks extend beyond memorization alone and include broader issues of data handling, inference-time exposure, and governance ambiguity \cite{Shanmugarasa2025PrivacyParadox, Yan2025PrivacyLLMs}. Moreover, in adjacent high-stakes domains such as health applications, researchers point out that the terms governing generic LLM services can be misaligned with the requirements of regulated use, creating additional legal and operational uncertainty \cite{Freyer2024HealthLLMs}. Therefore, in critically sensitive and privacy-concerning systems, the risk of proprietary LLM dependence lies less in what the provider's policy says today than in the fact that tomorrow's policy may be different.

To address this gap, the proposed system is designed around open-weight models, summary-based memory, structured state, and retrieval-based context for conversational scam analysis.

\section{System Design}

The proposed system advances an existing SMD, SmishX \cite{Wang2025SmishX}, into a system that also performs conversational scam detection (CSD). The development process is divided into three stages. In the first stage, closed-source dependencies are replaced due to privacy concerns. The second stage fixes major vulnerabilities of the pre-existing system, such as contact verification. The third stage adds a novel pipeline to detect conversational scams.

\subsection{Stage I: Privacy-Oriented Refactoring}
\label{stage1}

This stage replaces concerning external dependencies of the original system with open-source alternatives. In particular, the following changes were made:
\begin{itemize}

	\item \textbf{OpenAI API for LLM calls:} The baseline system uses OpenAI API for structured extraction of key information (brands and URLs) from SMS, summarizing page content, final phishing/spam classification, and to generate a user-friendly message. It also depends on OpenAI's vision model for screenshot understanding, interpreting webpage screenshots as an evidence stream for the final decision \cite{Wang2025SmishX, openai_vision_docs}. The proposed system replaces this with two open-weight models: deepseek-v3.1:671b for text and qwen3-vl:30b for vision \cite{deepseek_v3, qwen3_vl}. The proposed architecture, however, is not tied to these specific backends. Any sufficiently capable open-weight model exposing the same structured interface can be substituted. Section~\ref{subsec:csd_eval} (Table~\ref{tab:backend_portability}) verifies that the pipeline transfers to a second open-weight backend with comparable performance. The privacy benefit comes from the ability to run these models in a self-hosted configuration rather than from any single model choice.

	\item \textbf{Jina Reader API (HTML-to-text):} The base model uses Jina Reader \cite{jinareader2024} to extract plain text from HTML. The proposed system avoids third-party APIs for this task, since they can leak potentially sensitive indicators. For instance, the exact phishing link a user receives may contain unique tracking tokens tied to the recipient. It can also expose the user's browsing intent to the third party. In addition, contacting a URL through any third party can create an observable access event (e.g., server logs) that may increase the user's exposure to tracking or targeted follow-up. As such, Jina Reader is replaced with a local headless-browser crawler (built with Crawlee and Puppeteer) that renders the page, captures screenshots, and extracts visible/focused text, as shown in Figure~\ref{fig:jina-replace}. This makes the system more robust to JavaScript-heavy or visually deceptive phishing pages. Implementing a dedicated local PDF-processing pipeline is beyond the scope of this work; therefore, if a URL resolves to a PDF, the proposed system avoids standard webpage crawling and instead uses Docling. Docling, by IBM Research, is an open-source Python-based document-processing tool that extracts content into Markdown for downstream analysis \cite{ibm_docling_2024, docling_github}. The upgraded pipeline also performs early URL validation before launching the browser or network analysis. This prevents unnecessary visits to invalid URLs and reduces exposure to spurious artifacts.

	      \begin{figure}[H]
		      \centering
		      \includegraphics[width=\textwidth]{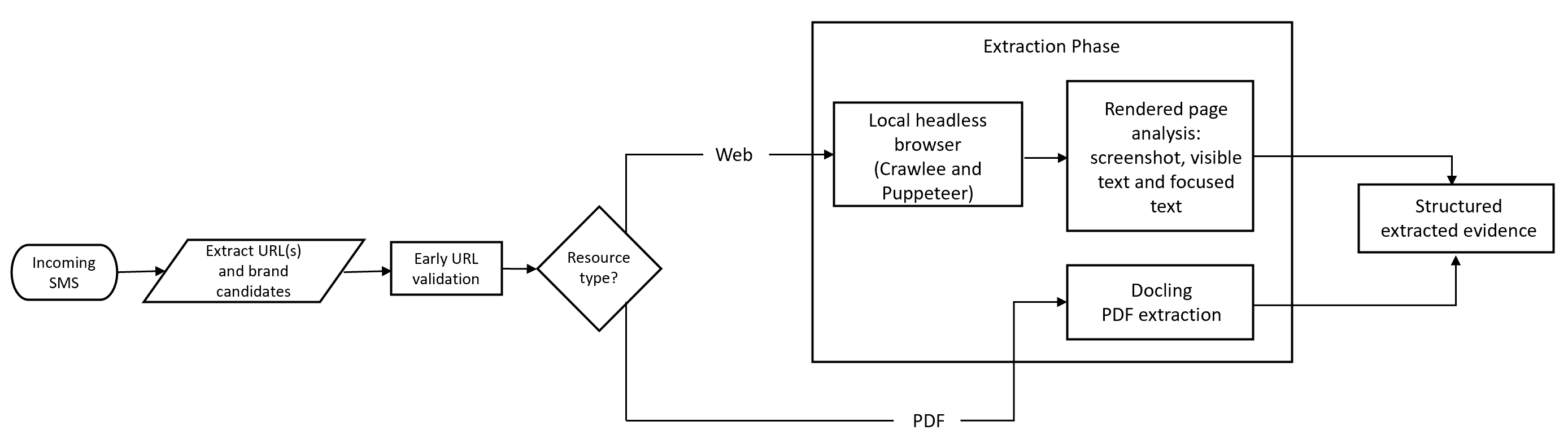}
		      \caption{The URL extraction and analysis pipeline. The system replaces third-party API dependencies with a local headless browser and Docling to prevent data leakage and tracking while improving the analysis of dynamic, JavaScript-heavy phishing content.}
		      \label{fig:jina-replace}
	      \end{figure}

	\item \textbf{Google Cloud Custom Search API \& Search Engine ID (brand-domain lookup):} The baseline model uses Google's Custom Search API (key + Search Engine ID) to query a brand name (e.g., ``Netflix'') and treats the top-ranked domain as the brand's ``true'' domain for mismatch checks. This structure imposed severe limitations. We replace it with Serper.dev, which provides Google-indexed search results through a stable API for the same ``brand-to-official-domain'' heuristic. Additionally, the base model is configured to draw only ``organic links'' from Google's Programmable Search Engine results, under the assumption that the top organic result contains the official site. This assumption is sometimes wrong, as Wikipedia, app stores, and directories can rank above the official site. As such, we additionally use Google's Knowledge Graph panel. For many well-known brands, Google recognizes the entity and attaches a Knowledge Graph panel, which Serper exposes as part of its JSON output on request \cite{google_kg_api,serper_google_search_api}. For our use case, that context is often a stronger ``official domain'' hint than whatever ranked \#1 organically, since it explicitly represents the entity's website rather than just a top-ranked page. We note that Serper.dev is a third-party service; this lookup, however, transmits only isolated brand names and extracted artifacts, not conversation content, user data, or any scam context. As such its privacy exposure is limited to verification queries rather than the sensitive dialogue itself. Achieving full data locality (no external calls at all) would instead require a self-hosted search index or a locally cached brand-to-domain allowlist, as noted in Section~\ref{sec:limitations}.

\end{itemize}

\subsection{Stage II: Strengthening Sender Authenticity}
\label{stage2}

\begin{figure}[h!]
	\centering
	\includegraphics[width=\textwidth]{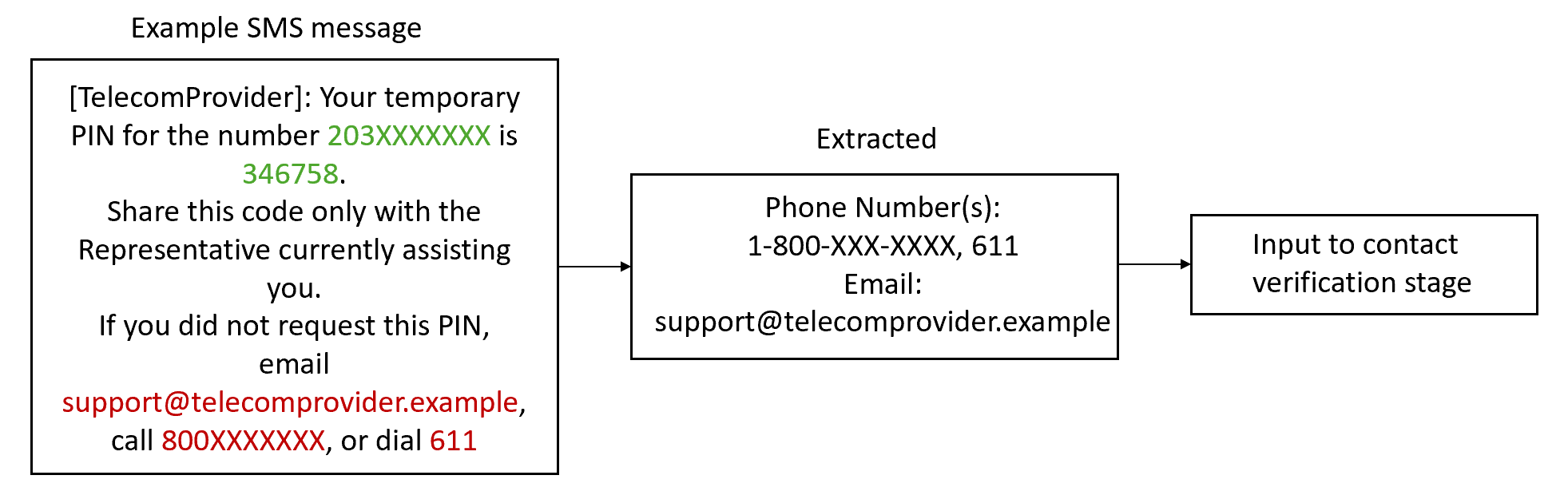}
	\caption{Example of contact information extraction from an SMS message. Both email addresses and phone numbers are identified and forwarded to the contact-verification stage.}
	\label{fig:contact-extract}
\end{figure}

The base model passes raw evidence from the extractor directly into the final detector, expecting it to infer sender authenticity on its own while deciding the legitimacy of the SMS. When tested, this configuration was unable to detect common phishing messages. Phishing SMS, in particular, use professional wording, familiar brand names, and legitimate links to the brand, while still containing fraudulent phone number(s) or email(s). In this stage, two major architectural modifications are made: the extraction pipeline is upgraded, and a verification agent is introduced before the final detector.
The extraction prompt enables the model to explicitly output phone numbers, email addresses, URLs, and brand names in a consistent, structured format, while preserving short codes in raw-digit form, as shown in Figure~\ref{fig:contact-extract}. The proposed pipeline instructs the LLM to extract only contact information intended to reach the sender, rather than arbitrary numeric strings in the message. Specific instructions are given not to extract contact information that does not belong to the sender (for instance, information belonging to the receiver/user), since the intention is solely to verify the sender's identity. This raw evidence is forwarded to the verification agent.

The verification stage is invoked conditionally rather than for every SMS. It runs only when the extraction stage identifies a claimed brand and sufficient verification targets (e.g., phone numbers, email addresses, or brand-related URL evidence). If these requirements are not met, the verification stage is skipped and the system proceeds directly to the final detection stage. This avoids unnecessary overhead when sender-authenticity checks are not applicable.

Through these significant modifications the proposed system now successfully verifies the sender's evidence while substantially reducing the load on the final detector. Instead of receiving raw search results, the final detection stage receives a compact verification summary. This makes it considerably easier for the final detector to decide the legitimacy of the SMS and reduces hallucination.

\subsection{Stage III: Conversational Scam Detection}
\label{stage3}

\begin{figure}
	\centering
	\includegraphics[width=0.86\linewidth]{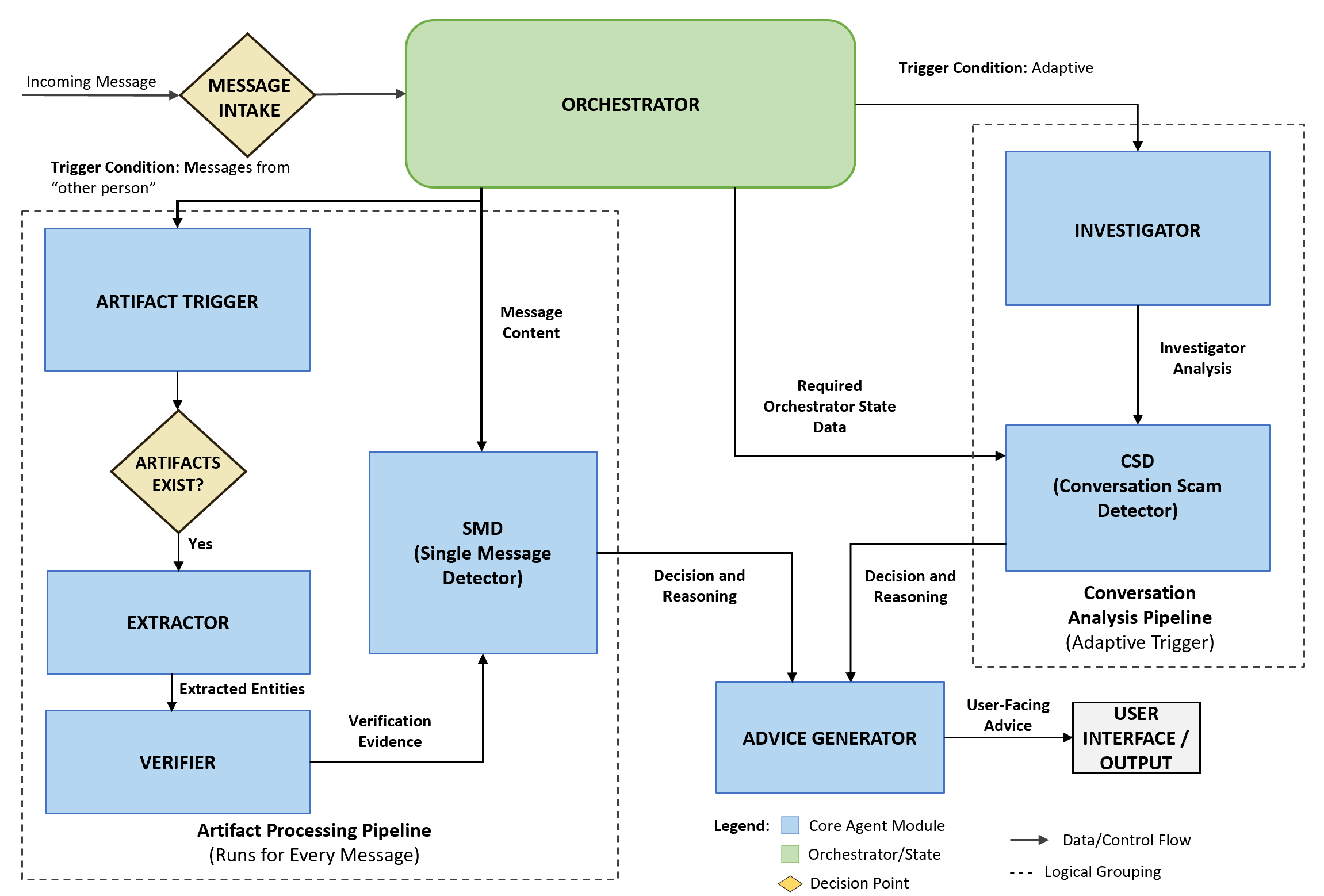}
	\caption{The proposed multi-agent system architecture. Messages are routed through an artifact-processing pipeline for single-message detection (SMD) and a conversation scam detector (CSD), with an advice generator producing user-facing output.}
	\label{fig:architecture}
\end{figure}

In this stage, a modular, multi-agent system capable of identifying conversational scams is developed. The system contains multiple specialized agents, where an ``orchestrator'' decides and coordinates the workflow for each agent in each run, as shown in Figure~\ref{fig:architecture}. The proposed system uses seven individual agents, described as follows.

\textbf{Artifact Trigger:} The purpose of this agent is only to detect artifacts. A lightweight prompt runs for every message from the ``sender'' and checks whether there exist channels (e.g., URL(s), phone number(s), and email(s)) to redirect the receiver/user. If artifacts are detected, the extractor agent is triggered.

\textbf{Extractor:} As described in Section~\ref{stage2}, the extractor agent extracts URL(s), phone number(s), email(s), and claimed brand(s) from the message and forwards them to the verifier.

\textbf{Verifier:} The verifier agent receives raw evidence and performs verification. The results are forwarded as a verification summary to the SMD module.

\textbf{SMD:} This agent takes the evidence from the verifier along with the original message and decides whether the message is scam, spam, or legitimate.

\textbf{Investigator:} This agent initially runs after the first three messages and produces a summary of the conversation so far along with its current threat level. The investigator inspects the ``claimed'' and ``inferred'' identity of the sender. The frequency of investigator runs is adaptive according to the current threat level. For instance, it runs every five messages if the threat level is none, and every two messages if it is high. The investigator then forwards this information to the CSD module.

\textbf{CSD:} This agent runs every time the investigator runs. It takes information from the investigator and the orchestrator. The orchestrator maintains information on recent messages, artifact events, and verification history. Based on the investigator's analysis and the orchestrator state, the CSD determines whether the conversation is leading to a scam.

\textbf{Advice Generator:} The user is not expected to be familiar with the model's technical details; to help build trust and improve accessibility, an advice-generator agent is developed. This agent creates short, simple, user-friendly advice based on raw evidences without overwhelming the users with technicality of the decision. Additionally, this agent supports multilingual output based on the user's language preference. Note that the full chain-of-thought is not performed in the user's preferred language, but in English, thus preserving the same underlying detection results regardless of language preference. Only the advice generator independently considers the user's choice of language. This substantially improves accessibility while maintaining scam-detection consistency. Overall, the advice generator is a critical interface layer between the analytical components of the system and the end user. Its main purpose is to ensure that the system is practically useful and not merely diagnostic.

The orchestrator makes three particularly important decisions. First is the rationale for running the artifact trigger only for messages from the ``Sender.'' Rather than running the trigger agent for all messages, it runs only if the message originates from an external party. This restriction reduces unnecessary computation and avoids false positives on benign user-authored messages. Second, instead of providing full conversation history to the CSD, the investigator supplies an incrementally updated summary. LLMs have a limited context window and are prone to hallucination on long prompts. Therefore, the summary makes the system stable and scalable for longer conversations, reduces unnecessary context growth, and helps CSD focus on the most relevant suspicious developments. Third, the investigator and CSD are instructed not to execute on every turn, but instead on an adaptive schedule. This is because conversational scam signals are expected to emerge over multiple turns rather than within isolated messages. Isolated harmful messages are expected to be handled by the SMD instead. The CSD first runs after three messages, which is considered the minimum number of messages to be treated as a conversation in this work. The investigator keeps track of the threat level, and if it increases, the analysis frequency is raised accordingly. On the other hand, if the threat level is none, the analysis frequency relaxes to every five turns. This allows the system to adapt and respond more aggressively to suspicious interactions.

\subsection{Testbed Interface}

A lightweight frontend testbed was developed using Streamlit \cite{streamlit_docs} for interactive testing of the system. The purpose of this interface is to provide a controlled environment in which the behavior of the proposed multi-agent system can be observed. Since the system is designed to detect scams that happen over multiple turns, the testbed provides a simple interface for entering messages turn by turn.

The interface is kept straightforward and functions as a two-party conversation simulator where messages can be entered from either the \textit{User} side or the \textit{Sender/Other Person} side (who may or may not be a scammer). Each message is forwarded to the backend orchestrator, which updates the internal conversation state and determines the execution of specialized agents (artifact trigger, extraction, verification, SMD, investigator updates, conversation-level detection, and advice generation). The testbed allows complete end-to-end workflow to be observed in an interactive setting, where practical behaviors such as incremental conversations across multiple messages can be tested. The frontend also includes a dropdown for selecting the advice output language. It allows the observer to inspect when warnings are triggered and to analyze whether a detection arises from single-message evidence or from broader conversational patterns. The interface, therefore, is convenient for both debugging and qualitative system evaluation.

\section{Datasets}

\subsection{Dataset for SMD}

The proposed SMD from Stage~II (Section~\ref{stage2}) is evaluated using the dataset provided by the authors of the baseline work \cite{Wang2025SmishX}. It contains real-world messages that were manually labeled into three categories: legitimate, phishing, and spam. Before evaluation, an independent internal audit was performed on the dataset. Out of the 1,200 messages, 636 (53\%) contain URL links, and 20 messages were found to be incorrectly labeled (15 legitimate and 3 spam messages that were in fact phishing, and 2 legitimate messages that were spam). Further investigation also revealed that some messages contained unverifiable or malformed URLs but were still marked as legitimate. A few obvious spam messages were likewise incorrectly labeled as legitimate. One such incorrectly labeled message is shown in Figure~\ref{fig:online-transport-spam-example}. 

Additionally, only limited coverage of scam messages containing phone number(s) and/or email(s) was found in the dataset. Therefore, 10 more such messages were collected from the internet and added. The contacts were replaced with realistic fabricated alternatives for privacy preservation and to avoid exposing real personal contact information in the dataset. Overall, the dataset was corrected, re-labeled, and augmented. After these changes, the dataset contains 287 phishing, 318 spam, and 605 legitimate messages (1,210 total) and is used for evaluating the proposed SMD.

\begin{figure}[!htb]
	\centering
	\fbox{%
		\begin{minipage}{0.92\linewidth}
			\small ``Online Transport wants to speak w/ you about a driving position! Click here [URL] to complete a full application or call [phone number] for more info. STOP to end''
		\end{minipage}%
	}
	\caption{An example message labeled as legitimate in the baseline dataset. However, its recruitment-style content, shortened URL, and bulk-SMS opt-out phrase make it more consistent with a spam message.}
	\label{fig:online-transport-spam-example}
\end{figure}

\subsection{Datasets for CSD}

At the time this study was conducted, no established open benchmark existed for multi-turn conversational scam detection. LoveFraud02 \cite{Faber2024LoveFraud02}, the closest available resource, is a single-family romance-fraud corpus. It had to be normalized into a common schema and truncated to its pre-scam portion before it could serve as a detection benchmark.  Alongside the proposed detection architecture, this work addresses the scarcity of such dataset. We build and release ConScamBench-278, a multi-category benchmark for multi-turn conversational scam detection covering eight distinct scam families. Concurrent work characterizes the structure of conversational scams without supplying a detector \cite{luo2026anatomy}. What sets the present system apart is that it runs fully self-hosted, unifying single-message and conversation-level detection under a bounded summary-based memory and produces user-facing explanations. The benchmark lets us evaluate the architecture across a broad range of scam types, and it gives later conversational detectors something concrete to be compared against.

Two datasets used for the evaluation of the CSD are LoveFraud02 \cite{Faber2024LoveFraud02} and our own curated dataset, ConScamBench-278.

LoveFraud02 \cite{Faber2024LoveFraud02} corpus consists of 83 real conversations (51,936 messages overall in the original release, spanning up to 3,345 turns) and was collected between 2021 and 2025. It was compiled to support linguistic and discourse analysis of romance-fraud scripts. It contains conversations between the author and fraudsters impersonating military officers, doctors, engineers, oil workers, businessmen, and celebrities. We normalize the dataset into a JSON schema in which each conversation is a multi-turn dialogue between a victim-facing actor (``Sender/Other Person'') and the recipient (``User''). To evaluate detection under realistic deployment conditions, we truncate each conversation to its \emph{pre-scam} portion and drop messages that occur only after the scam has been consummated, such as the victim's later realization and distress or the fraudster's eventual admission. A useful detector has to flag the fraud \emph{before} it succeeds, not recognize its aftermath. Operationally, the last retained turn is the final turn before the fraud is consummated or acknowledged. We keep the grooming phase, the scam solicitation and cut only post-consummation turns (a completed transfer, the victim's realization, or the fraudster's admission). Here, two considerations bound the resulting label-leakage risk. First, since the retained portion still contains the scam attempt, the reported recall measures detection of conversations that reach an overt solicitation, not pre-solicitation prediction. Second, and more importantly, detection happens early relative to the retained length: a median of 6 turns (mean 9.06) against a median conversation length of 469 turns. The system flags these conversations during grooming, long before the solicitation that ends the retained portion. So, recall is not an artifact of a solicitation sitting at the truncation boundary. The truncated corpus keeps all 83 conversations (47,452 messages) and remains long and realistic, spanning 11 to 2,622 turns with a median of 469 and a mean of 571.7 messages per conversation. What corpus lacks is variety in scam type, so a second dataset was needed. We, therefore, curate ConScamBench-278 that covers eight scam categories: gift-card, imposter, family, job-offer, pig-butchering, romance, task, and wrong-number scams.

The scam conversations combine real-world fraud with LLM-generated adversarial simulations. Of the 128 scam conversations, 110 are real and curated from public scam reports and shared anti-fraud archives. The remaining 18 were written by LLM (a small set per category) to cover the AI-assisted threat model that motivates this work, in which scammers use LLMs to automate and personalize persuasion at scale. Table \ref{tab:conscambench_summary} gives the per-category real/synthetic split. All real conversations were de-identified before inclusion, to protect victim privacy and to avoid republishing live scam infrastructure. Personal names and contact artifacts (phone numbers, email addresses, links, and cryptocurrency wallet addresses) were replaced with fictional placeholders, while the scam behavior and any impersonated brand names were kept. The dataset also contains 150 non-scam conversations. Of these, 135 come from publicly available, licensed human--human dialogue corpora (DailyDialog~\cite{li-etal-2017-dailydialog}, MultiWOZ~\cite{budzianowski-etal-2018-multiwoz}, EmpatheticDialogues~\cite{rashkin-etal-2019-towards}, and Deal-or-No-Deal~\cite{lewis-etal-2017-deal}), and 15 are hard negatives we wrote ourselves. This includes situations that look scam-adjacent but resolve benignly, such as a legitimate bank fraud-alert exchange or a colleague's gift-card request verified through normal channels, etc. The benign set is large enough to estimate false-positive behavior with reasonable precision. Every conversation was reviewed and labeled by hand before inclusion. For label reliability, a second annotator (a cybersecurity researcher who is not an author) independently double-coded a stratified subset of 80 conversations, eight from each scam category and sixteen benign. Agreement was 90.0\% (72/80), with Cohen's $\kappa = 0.89$ computed over the nine-way label assignment (eight scam categories plus benign) rather than within individual category strata. The eight disagreements were settled by consensus before release. Table~\ref{tab:inter_annotator_agreement} gives per-category counts. We report $\kappa$ only overall, since the per-category strata are too small for stable category-specific estimates.

\begin{table}[H]
	\centering
	\footnotesize
	\setlength{\tabcolsep}{3pt}
	\renewcommand{\arraystretch}{1.08}
	\caption{Inter-annotator agreement for ConScamBench-278 on a stratified double-coded subset. Category-level rows report disagreement counts only; Cohen's $\kappa$ is reported only overall because per-category strata are too small for stable category-specific $\kappa$ estimates.}
	\label{tab:inter_annotator_agreement}
	\begin{tabularx}{\textwidth}{>{\raggedright\arraybackslash}p{3.25cm} >{\centering\arraybackslash}p{1.75cm} >{\centering\arraybackslash}p{1.55cm} >{\centering\arraybackslash}p{1.85cm} X}
		\toprule
		\textbf{Category}                 & \textbf{\# Conversations} & \textbf{\# Annotators} & \textbf{\# Disagreements} & \textbf{Resolution Method}                                                       \\
		\midrule
		Gift card scam                    & 8                         & 2                      & 0                         & No adjudication required                                                         \\
		Imposter scam          & 8                         & 2                      & 1                         & Resolved by discussion using the scam-type definition                            \\
		Family scam & 8                         & 2                      & 1                         & Resolved by discussion using the scam-type definition                            \\
		Job offer scam                    & 8                         & 2                      & 1                         & Resolved by discussion using the scam-type definition                            \\
		Pig-butchering scam               & 8                         & 2                      & 1                         & Resolved by discussion using the scam-type definition                            \\
		Romance scam                      & 8                         & 2                      & 2                         & Resolved by discussion using the scam-type definition                            \\
		Task scam                         & 8                         & 2                      & 1                         & Resolved by discussion using the scam-type definition                            \\
		Wrong-number scam                 & 8                         & 2                      & 1                         & Resolved by discussion using the scam-type definition                            \\
		Non-scam / benign                 & 16                        & 2                      & 0                         & No adjudication required                                                         \\
		\midrule
		\textbf{Overall}                  & \textbf{80}               & \textbf{2}             & \textbf{8}                & \textbf{Overall Cohen's $\kappa = 0.89$; disagreements adjudicated by consensus} \\
		\bottomrule
	\end{tabularx}
\end{table}

\begin{table}[H]
	\centering
	\small
	\caption{Summary of ConScamBench-278 by conversation category. Real/Synth.\ give the real-sourced versus LLM-generated split of the scam conversations.}
	\label{tab:conscambench_summary}
	\setlength{\tabcolsep}{3.5pt}
	\begin{tabular}{lccccccc}
		\toprule
		\textbf{Category} & \textbf{Scam} & \textbf{Real} & \textbf{Synth.} & \textbf{Non-scam} & \textbf{Total} & \textbf{Turn range} & \textbf{Median} \\
		\midrule
		Gift card scam & 15 & 12 & 3 & 0 & 15 & 7--34 & 14 \\
		Imposter scam & 20 & 17 & 3 & 0 & 20 & 4--71 & 16.5 \\
		Family scam & 16 & 15 & 1 & 0 & 16 & 7--30 & 15 \\
		Job offer scam & 16 & 13 & 3 & 0 & 16 & 8--37 & 14 \\
		Pig-butchering scam & 17 & 15 & 2 & 0 & 17 & 7--40 & 17 \\
		Romance scam & 12 & 10 & 2 & 0 & 12 & 5--63 & 14 \\
		Task scam & 15 & 14 & 1 & 0 & 15 & 7--58 & 16 \\
		Wrong-number scam & 17 & 14 & 3 & 0 & 17 & 5--37 & 14 \\
		Benign/non-scam conversations & 0 & -- & -- & 150 & 150 & 4--32 & 8 \\
		\midrule
		\textbf{Total} & \textbf{128} & \textbf{110} & \textbf{18} & \textbf{150} & \textbf{278} & \textbf{4--71} & \textbf{11} \\
		\bottomrule
	\end{tabular}
\end{table}

\paragraph{Data provenance, labeling, and ethics.} The real scam conversations come from publicly available internet sources on which the exchanges had been openly posted: scam-report communities, anti-fraud archives, and scam-baiting forums. We verified the provenance of each conversation against its original public source, and used no private, restricted, or agreement-bound data. A conversation was labeled a scam only if it met a three-part protocol. First, it was identified as a scam at its source. Second, it matched a documented scam playbook for its category. And, third, two independent cybersecurity researchers (not authors of this paper) confirm and assign its scam category (inter-annotator agreement $\kappa = 0.89$, reported above). Since conversations are genuine exchanges, we kept them close to verbatim instead of paraphrasing. Beyond the contact-artifact and personal-name redaction described above, we removed or generalized quasi-identifiers such as specific locations and dates where they appeared. A re-identification check on the de-identified conversations found them hard to trace back to any individual, though no anonymization guarantees perfect untraceability. The 18 LLM-generated conversations were produced with ChatGPT~5.4 as challenging adversarial cases. Every conversation was reviewed, real or synthetic, before inclusion. The benchmark is built entirely from public data with identifying information removed, so it poses minimal privacy risk. ConScamBench-278 will be released publicly upon publication of this paper.

Today, conversational scams are increasingly shaped by generative AI. As such, ConScamBench-278 includes LLM-augmented scam simulations alongside real-world scam conversations. The design choice reflects an adversarial setting in which scammers can use LLMs to automate persuasive messages, sustain longer conversations, personalize emotional manipulation, and hold back overt malicious requests until trust has been built. We do not treat the LLM-generated conversations as naturally occurring scam transcripts. Instead, they serve as controlled adversarial simulations, testing whether a detector can recognize the scam intent across longer, more coherent, and more manipulative multi-turn interactions.

To limit unrealistic or repetitive synthetic artifacts, each LLM-augmented conversation was reviewed, edited where necessary, and labeled before inclusion. Using an LLM both to generate part of the benchmark and (in the SMD comparison) as a detection backend does introduce a possible source of bias. We discuss this in Section~\ref{sec:limitations} and mitigate it by pairing ConScamBench-278 with the naturally occurring LoveFraud02 corpus, so that the evaluation covers both real-world conversational fraud and LLM-assisted scam scenarios.

\section{Results}

\subsection{Evaluation Protocol}

The evaluations test two complementary capabilities. First, the single-message detector (SMD) is evaluated on an individual SMS-style messages. This evaluation tests if the system can tell phishing, spam, and legitimate messages apart when suspicious artifacts such as URLs, phone numbers, email addresses, or brand claims appear in isolation. Second, the conversation detector (CSD) is evaluated on multi-turn conversations. This evaluation tests if the system can catch scams that emerge gradually through trust-building, repeated persuasion, redirection, or delayed financial requests. As the system is designed for two distinct detection settings, the evaluation follows a task-level validation strategy. The SMD runs on isolated SMS-style messages and is compared against SmishX, the closest single-message phishing baseline. The CSD is then evaluated on LoveFraud02 \cite{Faber2024LoveFraud02} and ConScamBench-278 under multi-turn scam conditions.

No established open detector exists for multi-turn conversational scam detection. Prior agentic detectors such as SmishX work on single messages and cannot consume multi-turn context. Instead of comparing against an off-the-shelf competitor, we anchor the CSD in three ways and set out the design space it occupies. First, single-message detection is represented by SmishX and, inside our own pipeline, by the SMD path. The component-attribution analysis in Table~\ref{tab:detection_source_distribution} counts how often conversation-level reasoning rather than single-message detection is responsible for a catch (94 of 126 true positives), a direct measure of what the CSD adds. Second, unassisted human performance is measured in User Study~A. Third, we consider the two natural alternatives for conversation memory. At one extreme, \emph{full-history} prompting re-supplies the whole transcript at each decision. It retains every long-range signal, but the context grows without limit, its cost is quadratic in the number of turns, and it eventually exceeds the model's context window; Section~\ref{sec:scalability} quantifies this. At the other extreme, a fixed \emph{recent-window} classifier bounds the context by using only the final $k$-message window. It discards earlier conversational state, so it cannot combine evidence that is incriminating only in combination across distant turns: an identity claim, promise, or contradiction made early and exploited much later, where the final local window on its own may say little. Summary-based memory sits between the two. Its per-decision context is bounded, as in a recent window, and it still carries a compressed, evolving record of earlier evidence, as in full history. Section~\ref{sec:baselines} reports the recent-window classifier as an operational bounded-context baseline, and Section~\ref{sec:scalability} measures the cost of full-history prompting. We release ConScamBench-278 so that later automated conversational detectors can be compared head-to-head with ours. The split mirrors how the system is meant to be deployed: the SMD handles artifact-rich single messages, the CSD handles longitudinal scam patterns.

\paragraph{Development and test separation.} To guard against inadvertent tuning on the evaluation data, we froze all agent prompts, structured-output schemas, decision thresholds, verification rules, and the adaptive scheduling policy before the final evaluation was run. Development and prompt design used only illustrative examples and the interactive testbed. The LoveFraud02 and ConScamBench-278 evaluation conversations were held out; we did not consult them when writing prompts, selecting thresholds, or choosing the schedule. What we report below is a single pass of the frozen system over conversations it had not seen, not the endpoint of iterative optimization against a test set.

\paragraph{Statistical analysis.} Proportions are reported with 95\% Wilson score confidence intervals. For the pre/post comparisons in User Study~B we use one-sided Wilcoxon signed-rank test, which suits ordinal Likert responses, and report the matched-pairs rank-biserial correlation as an effect size \cite{wilcoxon1992individual}. We assess significance at $\alpha=0.05$.

\subsection{SMD evaluation}
\label{subsec:smd_eval}

To compare against SmishX \cite{Wang2025SmishX}, we re-evaluate the SmishX baseline on the \emph{same} backend as our system, the open-weight deepseek-v3.1:671b (Section~\ref{stage1}). Holding the backend fixed isolates our architectural changes, the strengthened extractor and the verification agent, from the effect of the base model. SmishX survives the backend change well: its per-class recall shifts only slightly from its ChatGPT-5.4 configuration (phishing 92.6\% to 91.3\%, spam 99.1\% to 97.8\%, legitimate 97.9\% to 97.2\%), so the swap does not put the baseline at a disadvantage. Table~\ref{tab:smishx_vs_proposed_classwise} gives the per-class recall for both systems on deepseek-v3.1:671b. The proposed SMD raises phishing recall from 91.3\% to 100.0\% and legitimate recall from 97.2\% to 98.5\%, and loses a little spam recall, from 97.8\% to 96.9\%. One worry is that the perfect phishing recall is an artifact of our re-audit. To check, we recompute performance on only the messages whose labels we left unchanged, those that already agreed with the original SmishX labels (259 of the 287 phishing messages). Phishing recall stays at 100\% (259/259); spam recall (96.8\%) and the legitimate false-positive rate (1.5\%) match the full-set values almost exactly. The relabeling does not inflate the headline number.

\begin{table}[H]
	\centering
	\caption{Per-class detection rate (recall) for SmishX and the proposed system, both run on deepseek-v3.1:671b, on the audited SMS dataset.}
	\label{tab:smishx_vs_proposed_classwise}
	\begin{tabular}{lcc}
		\toprule
		\textbf{Class} & \textbf{SmishX} & \textbf{Proposed System} \\
		\midrule
		Phishing       & 91.3\%          & 100.0\%                  \\
		Spam           & 97.8\%          & 96.9\%                   \\
		Legitimate     & 97.2\%          & 98.5\%                   \\
		\bottomrule
	\end{tabular}
\end{table}

Spam is where the proposed system does comparatively worse. A plausible cause is its heightened sensitivity to features that spam and legitimate messages share. Spam messages from well-known brands, carrying legitimate links and contacts, were often judged legitimate by the proposed SMD. Looking into these errors, the URLs usually passed the verifier, and the detector then read the whole message as legitimate. A few legitimate messages flagged as phishing would have been classified correctly with more context; one entry in the dataset is a prank whose nature only becomes clear in context. Stage~III of this study addresses that gap. The trade-off aside, overall the results point to a more robust framework. One further observation is that SmishX caught phishing when the fraudulent contacts belonged to well-known brands but missed it for lesser-known ones, which is exactly what the proposed verifier addresses.

\subsection{CSD evaluation}
\label{subsec:csd_eval}

Table~\ref{tab:lovefraud02_summary} gives the performance of the CSD on the public LoveFraud02 dataset.
The proposed system detects all 83 conversational scams (recall 100\%, 95\% CI [95.6, 100]), with a mean first-detection latency of 9.06 turns and a median of 6. First-detection latency is computed over detected conversations only, on the truncated pre-scam corpus.

\begin{table}[H]
	\centering
	\caption{Summary of scam-detection results on the LoveFraud02 dataset.}
	\label{tab:lovefraud02_summary}
	\begin{tabular}{lc}
		\toprule
		\textbf{Metric}              & \textbf{Value}                  \\
		\midrule
		Total Conversations          & 83                              \\
		Scam Detected                & 83 (100\%, 95\% CI [95.6, 100]) \\
		Average First Detection Turn & 9.06                            \\
		Median First Detection Turn  & 6                               \\
		\bottomrule
	\end{tabular}
\end{table}

\begin{table}[H]
	\centering
	\caption{Confusion matrix from the evaluation on ConScamBench-278.}
	\label{tab:curated_confusion_matrix}
	\begin{tabular}{lcc}
		\toprule
		\textbf{}                  & \textbf{Predicted Scam} & \textbf{Predicted Legitimate} \\
		\midrule
		\textbf{Actual Scam} & 126 & 2 \\
		\textbf{Actual Legitimate} & 4 & 146 \\
		\bottomrule
	\end{tabular}
\end{table}

\begin{table}[H]
	\centering
	\small
	\caption{Conversation-level performance on ConScamBench-278, with 95\% Wilson confidence intervals, reported on the full benchmark and on the real-sourced scam subset (excluding the 18 LLM-generated adversarial simulations).}
	\label{tab:conscambench_metrics}
		\begin{tabular}{lcc}
			\toprule
			\textbf{Metric} & \textbf{Full benchmark} & \textbf{Real-sourced only} \\
			 & \textbf{(128 scam)} & \textbf{(110 scam)} \\
			\midrule
			Accuracy & 97.8\% [95.4, 99.0] & 97.7\% [95.1, 98.9] \\
			Scam precision & 96.9\% [92.4, 98.8] & 96.4\% [91.2, 98.6] \\
			Scam recall & 98.4\% [94.5, 99.6] & 98.2\% [93.6, 99.5] \\
			Scam F1-score & 97.7\% & 97.3\% \\
			Specificity (non-scam) & 97.3\% [93.3, 99.0] & 97.3\% [93.3, 99.0] \\
			False-positive rate & 2.7\% [1.0, 6.7] & 2.7\% [1.0, 6.7] \\
			False-negative rate & 1.6\% [0.4, 5.5] & 1.8\% [0.5, 6.4] \\
			\bottomrule
		\end{tabular}
\end{table}

Tables~\ref{tab:curated_confusion_matrix} and~\ref{tab:conscambench_metrics} show results on our curated dataset, ConScamBench-278. The system detects 126 of the 128 scam conversations (recall $98.4\%$, 95\% CI [94.5, 99.6]). Both misses are real-sourced conversations that sit on the boundary. One is a pig-butchering exchange that reads like a familiar interpersonal conversation and never reaches a concrete scam action, no investment, payment, link, secrecy request, or credential disclosure. The other is a wrong-number exchange that ends after a brief setup and a business-partner excuse, before any friendship pivot, financial request, artifact, or off-platform redirection that would justify a scam label. Every one of the 18 LLM-generated adversarial simulations was detected. Restricting the metrics to the 110 real-sourced scams as seen in Table~\ref{tab:conscambench_metrics}, right column, barely moves them with recall 98.2\% (95\% CI [93.6, 99.5]), precision 96.4\%, and the same 2.7\% false-positive rate. The headline result is not propped up by same-family synthetic data. Part of the benchmark is nonetheless LLM-assisted and evaluated with an LLM backend, so we read these figures as a strong controlled result and not as a direct estimate of real-world recall, whereas, the LoveFraud02 results discussed before (100\% recall) supply the naturally occurring counterpart. The detector also raises four false positives among the 150 non-scam conversations with a false-positive rate of 2.7\% (95\% CI [1.0, 6.7]). The benign conversations come from generic human-to-human dialogue corpora. As such, the rate describes behavior on real but not scam-adjacent chats. Neither corpus measures the deployment-critical quantity, false alarms on benign conversations arriving through the same channels as scams. LoveFraud02 contains no benign conversations at all. Section~\ref{sec:stress_test} stress tests that case directly with 76 authored benign conversations spanning the eight scam-adjacent families. All four false positives here were wrong-number interactions in which the initiating party dragged the exchange out, a pattern that closely resembles conversational scam behavior.

Table~\ref{tab:detection_source_distribution} counts how often each detection module contributed to a true-positive decision. The attribution works as a lightweight component analysis where instead of disabling modules in configurations that do not match the intended pipeline, it shows how often each detection path is responsible for a catch. The two modules cover different ground. The SMD catches suspicious standalone messages that redirect the user to harmful external channels; the CSD catches scam patterns that only surface over the course of a conversation. Of the 126 true positives, 94 (74.6\%) came from the conversation-level detector and 32 (25.4\%) from the single-message detector. Most of the catches, then, are ones a single-message detector on its own would miss.

Table~\ref{tab:smd_csd_overlap_buckets} makes the relationship between the two paths explicit. On ConScamBench-278 the single-message path adds no unique conversational recall. All 32 scams it detects in isolation are also detected by the conversation-level detector alone. None are detected only by the single-message detector. 94 are detected only by the conversation-level detector, whereas, 2 are detected by neither. As seen in Table~\ref{tab:smd_csd_latency_comparison}, averaged over detected scams, the conversation-level detector alone fires marginally earlier than the single-message detector alone (mean turn 5.08 versus 8.22), and the full pipeline averages to 5.17 detection turns. The single-message path contributes to the coverage of artifact-bearing single messages and first-trigger catches inside the pipeline (Table~\ref{tab:detection_source_distribution}), not extra conversational recall.

\begin{table}[H]
\centering
\caption{Detector-overlap buckets for scam conversations on ConScamBench-278, from the SMD-only and CSD-only ablations.}
\label{tab:smd_csd_overlap_buckets}
\begin{tabular}{lcc}
\toprule
\textbf{Bucket} & \textbf{Count} & \textbf{Share of scams} \\
\midrule
Detected by both SMD-only and CSD-only & 32 & 25.0\% \\
Detected only by SMD-only & 0 & 0.0\% \\
Detected only by CSD-only & 94 & 73.4\% \\
Detected by neither & 2 & 1.6\% \\
\bottomrule
\end{tabular}
\end{table}

\begin{table}[H]
\centering
\caption{Detection-turn comparison for the SMD-only and CSD-only ablations on ConScamBench-278. Averages are computed over detected scam conversations only.}
\label{tab:smd_csd_latency_comparison}
\begin{tabular}{lcc}
\toprule
\textbf{Configuration} & \textbf{Detected scams} & \textbf{Avg. detection turn} \\
\midrule
Full system & 126 & 5.17 \\
SMD-only & 32 & 8.22 \\
CSD-only & 126 & 5.08 \\
\bottomrule
\end{tabular}
\end{table}

\begin{table}[H]
	\centering
	\caption{First-trigger attribution for true-positive scam detections on ConScamBench-278 (which module fired first, not a counterfactual unique-contribution analysis).}
	\label{tab:detection_source_distribution}
	\begin{tabular}{lc}
		\toprule
		\textbf{Source}              & \textbf{Count} \\
		\midrule
		Conversation Scam Detector & 94 \\
		Single-message Scam Detector & 32 \\
		\bottomrule
	\end{tabular}
\end{table}

\paragraph{Backend portability.} To find out if the proposed pipeline depend on one specific backend, we re-ran the full CSD with the text backend swapped from deepseek-v3.1:671b to Qwen2.5-72B-Instruct. The vision backend, prompts, orchestration, decision cadence, and labels were left untouched. Table \ref{tab:backend_portability} shows the architecture transferring to the second open-weight backend with comparable performance. Accuracy and scam recall on ConScamBench-278 stay within one to two points of the primary configuration (97.1\% vs.\ 97.8\% accuracy; 96.9\% vs.\ 98.4\% recall). On LoveFraud02, scam recall holds at 98.8\%, and average detection comes somewhat later (turn 15.42 vs.\ 9.06). The detection behavior, therefore, belongs to the architecture rather than to one model. Although, the stronger backend does detect earlier and a little more completely.

\begin{table}[H]
	\centering
	\footnotesize
	\setlength{\tabcolsep}{3pt}
	\caption{Backend-portability check. The second backend uses the same CSD pipeline, prompts, labels, and decision cadence as the primary one.}
	\label{tab:backend_portability}
	\newcolumntype{Y}{>{\centering\arraybackslash}X}
	\begin{tabularx}{\textwidth}{>{\raggedright\arraybackslash}p{3.4cm}YYYY}
		\toprule
		\textbf{Backend}                      & \textbf{ConScamBench-278 Accuracy} & \textbf{ConScamBench-278 Scam Recall} & \textbf{LoveFraud02 Scam Recall} & \textbf{Avg.\ Detection Turn (LoveFraud02)} \\
		\midrule
		deepseek-v3.1:671b (main) & 97.8\% & 98.4\% & 100.0\% & 9.06 \\
		Qwen2.5-72B-Instruct (second backend) & 97.1\% & 96.9\% & 98.8\% & 15.42 \\
		\bottomrule
	\end{tabularx}
\end{table}

\subsection{Comparison with Conversational Baselines}
\label{sec:baselines}

We compare the full proposed pipeline against two baselines that share its data and backend LLM (deepseek-v3.1:671b). The first is a bounded-cost \emph{recent-window} classifier, which decides from the final $k$-message window alone. The second is a \emph{full-transcript} classifier, which receives the whole conversation in a single call. It is an offline, high-information baseline, since it sees everything at once instead of deciding online as messages arrive. Tables \ref{tab:conscambench_baselines} and \ref{tab:lovefraud_baselines} give the results on ConScamBench-278 and LoveFraud02.

\begin{table}[H]
	\centering\small
	\caption{Conversational baselines on ConScamBench-278 (128 scam, 150 non-scam), with the same backend LLM and the same ground truth. The recent-window classifier decides from the final window; the full-transcript classifier receives the whole conversation in one call.}
	\label{tab:conscambench_baselines}
	\begin{tabular}{lcccc}
		\toprule
		Method                  & Acc.   & Scam Prec. & Scam Rec. & FPR \\
		\midrule
		Recent-window ($k{=}6$) & 82.0\% & 100.0\% & 60.9\% & 0.0\% \\
		Full-transcript classifier & 97.8\% & 98.4\% & 96.9\% & 1.3\% \\
		Proposed System & 97.8\% & 96.9\% & 98.4\% & 2.7\% \\
		\bottomrule
	\end{tabular}
\end{table}

\begin{table}[H]
	\centering\small
	\caption{Conversational baselines on LoveFraud02 (83 scams; all-scam, so precision and FPR are uninformative).}
	\label{tab:lovefraud_baselines}
	\begin{tabular}{lc}
		\toprule
		Method                  & Scam Recall \\
		\midrule
		Recent-window ($k{=}6$) & 48.2\% \\
		Full-transcript classifier & 97.6\% \\
		Proposed System         & 100.0\% \\
		\bottomrule
	\end{tabular}
\end{table}

On longer conversations in LoveFraud02 (median 469 turns), the recent-window classifier finds only 48.2\% of scams (95\% CI [37.8, 58.8]) as the decisive evidence rarely sits in the final window. A final-window classifier also cannot warn anyone until the conversation ends, so it offers no early detection at all. The proposed system, carrying a running summary, finds all 83 scams and finds them early (mean turn 9.06). This is the empirical counterpart of the design argument discussed in Section~\ref{sec:scalability}. A bounded window with no memory loses exactly the long-range signal that summary-based memory keeps. On the shorter ConScamBench-278 conversations, the recent-window classifier does better (60.9\% recall, 95\% CI [52.3, 69.0] as shown in Table~\ref{tab:window_sweep}, with no false positives, but still misses nearly two in five scams. Sweeping the window size raises recall monotonically from 55.5\% at $k{=}4$ to 82.0\% at $k{=}16$, and even the largest window stays well below the proposed system's 98.4\% while costing a proportionally larger per-decision context. The bounded window trades context cost for recall without closing the gap, and since it decides only from the final window it still cannot warn early. Its precision stays at 100\% (no false positives) at every $k$.

\begin{table}[H]
\centering\small
\caption{Recent-window classifier on ConScamBench-278 as a function of window size $k$ (same backend and ground truth). Precision is 100\% and the false-positive rate 0.0\% at every $k$. The proposed system attains 98.4\% recall at a 2.7\% false-positive rate.}
\label{tab:window_sweep}
\begin{tabular}{ccc}
\toprule
\textbf{Window $k$} & \textbf{Scam recall (95\% CI)} & \textbf{Accuracy} \\
\midrule
4  & 55.5\% [46.8, 63.8] & 79.5\% \\
6  & 60.9\% [52.3, 69.0] & 82.0\% \\
8  & 66.4\% [57.9, 74.0] & 84.5\% \\
12 & 78.1\% [70.2, 84.4] & 89.9\% \\
16 & 82.0\% [74.5, 87.7] & 91.7\% \\
\bottomrule
\end{tabular}
\end{table}

The full-transcript classifier sees the entire conversation at once and still does not beat the proposed system. On ConScamBench-278, it misses four scams to our two (96.9\% vs.\ 98.4\% recall) and reaches the same overall accuracy (both 97.8\%), though with a lower false-positive rate (1.3\% vs.\ 2.7\%). On LoveFraud02, it misses two scams (97.6\% recall, 95\% CI [91.6, 99.3]) whereas the proposed system misses none. Language models are known to lose salient details in long or cluttered contexts \cite{laban2026lost}, and an entire multi-turn transcript can dilute the decisive scam signals that summary-based memory distills and keeps. Read together with the cost analysis of Section~\ref{sec:scalability}, this puts summary-based memory at least on par with full-transcript prompting and far cheaper. It matches the full-transcript classifier's accuracy on ConScamBench-278 and beats LoveFraud02. It catches more scams on both corpora and costs a fraction of the context. Its one disadvantage is a marginally higher false-positive rate.

\subsection{Non-Agentic and Reduced-State Baselines}
\label{sec:p4_baselines}

The recent-window and full-history comparisons above vary how much conversation context each decision sees. To isolate the contribution of the agentic structure itself, we compare the full pipeline against three reduced configurations on ConScamBench-278, holding the backend, ground truth, and decision rule fixed (Tables~\ref{tab:p4_baseline_comparison} and~\ref{tab:p4_baseline_ci}). The \emph{non-agentic LLM classifier} replaces the whole multi-agent pipeline with one monolithic classification call over the conversation, dropping the investigator, the verifier, the single-message path, and the orchestration. The \emph{summary-only} baseline keeps the running summary as memory but strips the surrounding agents. As such, the decisions come from the compressed summary alone, without artifact verification or single-message screening. The \emph{no-scheduling} baseline keeps the full agent set, disables the adaptive schedule and re-evaluates on a fixed cadence instead of escalating as the running threat level rises.

Each reduction degrades detection along an interpretable axis. The monolithic non-agentic classifier is weakest, at 94.5\% recall (95\% CI [89.1, 97.3]) and a 6.7\% false-positive rate (95\% CI [3.7, 11.8]). It over-flags benign conversations that mention gift cards, bank alerts, family emergencies, or wrong numbers, which are precisely the artifact- and pretext-driven confusions that the verifier and the single-message path exist to resolve. The summary-only baseline brings the false-positive rate down to 4.7\%, but its recall falls to 93.8\%: compressing the dialogue into a summary without the verification. Single-message agents loses fine-grained evidence such as delayed pivots, inconsistent identities, and artifact-specific cues. The no-scheduling baseline comes closest, at 96.9\% recall and a 2.0\% false-positive rate. Yet, disabling adaptive escalation lowers recall and delays detection (mean turn 6.42 against 5.17 for the full system) because scams that only become evident once new context accumulates are re-evaluated too rarely. The full system reaches the best recall (98.4\%, 95\% CI [94.5, 99.6]) and the earliest average detection (turn 5.17), at a false-positive rate (2.7\%) statistically indistinguishable from the reduced configurations. Table~\ref{tab:p4_baseline_error_patterns} summarizes the dominant error pattern of each configuration. Each reduction fails in the way its removed component was meant to prevent, while the full system's residual errors are confined to ambiguous short exchanges and to benign hard negatives we deliberately shaped to look like scams. The agentic structure and the adaptive schedule each contribute measurably, and no single-call or reduced-state substitute matches the full pipeline on recall and detection latency at a comparable false-positive rate.

\begin{table}[H]
\centering\small
\caption{Comparison with non-agentic and reduced-state baselines on ConScamBench-278. Every configuration runs on the same 128 scam and 150 legitimate conversations. Detection-turn averages cover detected scam conversations only.}
\label{tab:p4_baseline_comparison}
\resizebox{\textwidth}{!}{%
\begin{tabular}{lccccccccc}
\toprule
\textbf{Configuration} & \textbf{TP} & \textbf{FN} & \textbf{FP} & \textbf{TN} & \textbf{Accuracy} & \textbf{Precision} & \textbf{Scam Recall} & \textbf{FPR} & \textbf{Avg. Det. Turn} \\
\midrule
Non-agentic LLM classifier & 121 & 7 & 10 & 140 & 93.9\% & 92.4\% & 94.5\% & 6.7\% & 7.31 \\
Summary-only baseline & 120 & 8 & 7 & 143 & 94.6\% & 94.5\% & 93.8\% & 4.7\% & 7.88 \\
No-scheduling baseline & 124 & 4 & 3 & 147 & 97.5\% & 97.6\% & 96.9\% & 2.0\% & 6.42 \\
Proposed system (full) & 126 & 2 & 4 & 146 & 97.8\% & 96.9\% & 98.4\% & 2.7\% & 5.17 \\
\bottomrule
\end{tabular}}
\end{table}

\begin{table}[H]
\centering\small
\caption{Recall and false-positive confidence intervals for the reduced-state baselines on ConScamBench-278. Wilson 95\% confidence intervals are reported for scam recall and false-positive rate.}
\label{tab:p4_baseline_ci}
\begin{tabular}{lcccc}
\toprule
\textbf{Configuration} & \textbf{Scam Recall} & \textbf{95\% CI} & \textbf{FPR} & \textbf{95\% CI} \\
\midrule
Non-agentic LLM classifier & 94.5\% & [89.1, 97.3] & 6.7\% & [3.7, 11.8] \\
Summary-only baseline & 93.8\% & [88.2, 96.8] & 4.7\% & [2.3, 9.3] \\
No-scheduling baseline & 96.9\% & [92.2, 98.8] & 2.0\% & [0.7, 5.7] \\
Proposed system (full) & 98.4\% & [94.5, 99.6] & 2.7\% & [1.0, 6.7] \\
\bottomrule
\end{tabular}
\end{table}

\begin{table}[H]
\centering\small
\caption{Qualitative error patterns for the reduced-state baselines on ConScamBench-278.}
\label{tab:p4_baseline_error_patterns}
\begin{tabularx}{\textwidth}{>{\raggedright\arraybackslash}p{3.2cm} >{\centering\arraybackslash}p{1.5cm} >{\centering\arraybackslash}p{1.5cm} X}
\toprule
\textbf{Configuration} & \textbf{False negatives} & \textbf{False positives} & \textbf{Dominant error pattern} \\
\midrule
Non-agentic LLM classifier & 7 & 10 & Missed slow-burn or ambiguous scams without explicit payment artifacts; over-flagged benign conversations containing gift cards, bank alerts, family emergencies, or wrong-number cues. \\
Summary-only baseline & 8 & 7 & Lost fine-grained scam evidence during compression, especially delayed pivots, inconsistent identities, and artifact-specific cues. \\
No-scheduling baseline & 4 & 3 & Detected most obvious scams but missed or delayed cases that required repeated re-evaluation after new context accumulated. \\
Proposed system (full) & 2 & 4 & Remaining false negatives were short or socially familiar ambiguous exchanges; remaining false positives came from intentionally scam-shaped benign hard negatives. \\
\bottomrule
\end{tabularx}
\end{table}

\subsection{Component Ablation}
\label{sec:ablation}

We ablate the pipeline one element at a time, holding the backend (deepseek-v3.1:671b), the labels, and the decision prompts fixed and varying only the named component. Tables~\ref{tab:ablation_conscambench} and~\ref{tab:ablation_lovefraud} report the results on ConScamBench-278 and LoveFraud02. As throughout, detection-turn averages cover detected scam conversations only.

\begin{table}[H]
	\centering
	\caption{Ablation results on ConScamBench-278. Every row uses the same backend (deepseek-v3.1:671b), labels, and decision prompts; only the named component varies. Detection-turn averages cover detected scam conversations only.}
	\label{tab:ablation_conscambench}
	\begin{tabularx}{\textwidth}{lcccc}
		\toprule
		\textbf{Configuration}     & \textbf{Scam Recall} & \textbf{Precision} & \textbf{FPR} & \textbf{Avg. Detection Turn} \\
		\midrule
		Full system (baseline) & 98.4\% & 96.9\% & 2.7\% & 5.17 \\
		-- no verifier & 98.4\% & 92.6\% & 6.7\% & 5.03 \\
		-- no investigator/summary & 96.1\% & 94.6\% & 4.7\% & 7.34 \\
		SMD-only (CSD disabled) & 25.0\% & 97.0\% & 0.7\% & 8.16 \\
		CSD-only (SMD disabled) & 98.4\% & 97.7\% & 2.0\% & 5.08 \\
		\bottomrule
	\end{tabularx}
\end{table}

\begin{table}[H]
	\centering
	\caption{Ablation results on LoveFraud02. Since LoveFraud02 is a scam-only corpus here, we report only scam recall and first-detection timing. Detection-turn averages cover detected scam conversations only.}
	\label{tab:ablation_lovefraud}
	\begin{tabularx}{0.8\textwidth}{lcc}
		\toprule
		\textbf{Configuration}     & \textbf{Scam Recall} & \textbf{Avg. Detection Turn} \\
		\midrule
		Full system (baseline)     & 100.0\%              & 9.06                         \\
		-- no verifier             & 99.1\%               & 9.14                         \\
		-- no investigator/summary & 95.2\%               & 13.84                        \\
		SMD-only (CSD disabled)    & 7.1\%                & 312.75                       \\
		CSD-only (SMD disabled)    & 100.0\%              & 9.84                         \\
		\bottomrule
	\end{tabularx}
\end{table}

From these results, three patterns stand out. First, removing the investigator/summary lowers recall (98.4\% to 96.1\% on ConScamBench-278; 100.0\% to 95.2\% on LoveFraud02) and delays detection (turn 9.06 to 13.84 on LoveFraud02). The summary memory therefore helps catch scams, and helps catch them earlier. Section~\ref{sec:baselines} makes the stronger version of the point, where a memoryless recent-window baseline finds only 48.2\% of LoveFraud02 scams. The ablation study shows the summary component is part of what preserves long-range signal and the recent-window comparison shows how much is lost with no evolving memory at all. Second, the verifier mostly affects precision. Removing it leaves recall untouched and raises the false-positive rate (2.7\% to 6.7\% on ConScamBench-278). This is consistent with a component whose job is to verify sender authenticity rather than scam detection. Third, the SMD and CSD paths complement each other. On ConScamBench-278, the CSD-only configuration attains slightly higher precision (97.7\%) and a lower false-positive rate (2.0\%) than the full system, so the SMD path costs us one extra benign flag. We keep it anyway: it supplies the first-trigger detection for a quarter of the true positives in the full pipeline (25.4\%, as per Table~\ref{tab:detection_source_distribution}) and carries the entire single-message evaluation of Section~\ref{subsec:smd_eval}. The overlap analysis (Table~\ref{tab:smd_csd_overlap_buckets}) shows it adds no unique conversational recall here, so its role is coverage of artifact-bearing single messages, not extra recall or earlier detection. On LoveFraud02, the SMD-only configuration finds only 7.1\% of scams as conversational fraud of this kind is almost entirely a CSD task.

\subsection{Error Analysis}

On ConScamBench-278, the proposed system produced four false positives and two false negatives, a false-positive rate of 2.7\% (95\% CI [1.0, 6.7]) over 150 non-scam conversations. The pattern leans mildly toward false positives, four false alarms against two misses, which is what a cautious detector looks like. In scam prevention the trade may be acceptable, since missing a real scam can cost far more than warning a user about a suspicious but benign conversation. False positives still matter, although, too many warnings erode user trust and invite warning fatigue. All four false positives were benign wrong-number exchanges that ran unusually long.

Scam conversations in ConScamBench-278 are longer on average than benign ones (median 14 versus 8 turns). Table~\ref{tab:length_conditioned} experiments with several conversation length (turns). The four false positives are mid-length benign conversations of 6--9 turns, close to the benign median, not the longest benign conversations. Restricting the evaluation to the length range the two classes share (4--32 turns) leaves performance essentially where it was (recall 98.3\%, false-positive rate 2.7\%). Most directly, the system separates scam from benign within fixed length bands. On the shortest conversations ($\le 8$ turns) it reaches 96.3\% scam recall at a 3.2\% false-positive rate, and the separation holds in every band (Table~\ref{tab:length_conditioned}). Detection is driven by conversational content, not by length alone, although a residual length correlation remains and ruling it out fully would require length-matched data collection.

\begin{table}[H]
\centering\small
\caption{Detection performance by total conversation length on ConScamBench-278. Scam and benign stay separable in every length band, the shortest included, so the signal is not primarily length-driven.}
\label{tab:length_conditioned}
\begin{tabular}{lcccc}
\toprule
\textbf{Length (turns)} & \textbf{Scam n} & \textbf{Scam recall} & \textbf{Benign n} & \textbf{FPR} \\
\midrule
$\le 8$ & 27 & 96.3\% & 94 & 3.2\% \\
9--14 & 42 & 100.0\% & 20 & 5.0\% \\
$\ge 15$ & 59 & 98.3\% & 36 & 0.0\% \\
\bottomrule
\end{tabular}
\end{table}

\subsection{Scam-Adjacent Benign Stress Test}
\label{sec:stress_test}

The false-positive rate above is measured on generic human-to-human dialogue, which looks nothing like the channels through which scams actually arrive. To estimate false-alarm behavior on the deployment-critical case, i.e., benign conversations that superficially resemble scam entry points, we built a stress-test set of 76 scam-adjacent benign conversations. These conversations span the same eight situational families scammers exploit (benign wrong-number messages, legitimate recruiter outreach, genuine bank fraud-alert notifications, real family requests for help, ordinary marketplace haggling, authentic customer-support contacts, benign finance and crypto discussion, and verified gift-card coordination). Each conversation resolves benignly but opens on the pretext a scam would use. The set contains no contact artifacts by construction: no phone numbers, links, email addresses, or brand-impersonation strings. A detector cannot separate these from scams by keying on artifacts; the only signal available is conversational content and trajectory. The set is an adversarial probe of the CSD and not a naturally occurring corpus.

The system raises 5 false alarms out of 76 conversations, a false-positive rate of 6.6\% (95\% CI [2.8, 14.5]), reported in Table \ref{tab:scam_adjacent_benign_stress_test}. That exceeds the 2.7\% observed on generic benign dialogue, as expected, since these conversations are engineered to sit close to the scam decision boundary. Roughly nineteen in twenty scam-adjacent benign conversations are still left unflagged. The per-category breakdown (Table~\ref{tab:scam_adjacent_benign_by_category}) localizes the errors. The wrong-number family is once again the dominant failure mode (2 of 10, 20.0\%), as it was on ConScamBench-278, followed by one false alarm in family-emergency, finance/crypto and gift-card families. Recruiter, Bank-alert, marketplace, and support families produced none. The error taxonomy  in Table \ref{tab:scam_adjacent_benign_error_taxonomy} shows where the failures cluster. Benign exchanges whose surface trajectory, an unprompted stranger prolonging contact, a relative asking for money, a third party coordinating a gift-card purchase, mirrors an early scam script closely enough to trip the detector before the benign resolution arrives. Wrong-number continuation is the system's principal false-positive liability, and 6.6\% is a realistic upper bound on deployment false alarms from a controlled, authored probe. A benign corpus drawn from live scam-bearing channels would sharpen the estimate, discussed as limitations in Section~\ref{sec:limitations}.

\begin{table}[H]
\centering\small
\caption{Supplemental scam-adjacent benign stress test. All conversations are benign but resemble common scam channels or scam-adjacent situations.}
\label{tab:scam_adjacent_benign_stress_test}
\begin{tabular}{lcccc}
\toprule
\textbf{Set} & \textbf{Benign conversations} & \textbf{False positives} & \textbf{FPR} & \textbf{95\% CI} \\
\midrule
Scam-adjacent benign stress test & 76 & 5 & 6.6\% & [2.8, 14.5] \\
\bottomrule
\end{tabular}
\end{table}

\begin{table}[H]
\centering\small
\caption{False positives by scam-adjacent benign category.}
\label{tab:scam_adjacent_benign_by_category}
\begin{tabular}{lccc}
\toprule
\textbf{Category} & \textbf{Conversations} & \textbf{False positives} & \textbf{FPR} \\
\midrule
Wrong-number benign & 10 & 2 & 20.0\% \\
Recruiter benign & 9 & 0 & 0.0\% \\
Bank-alert benign & 10 & 0 & 0.0\% \\
Family-emergency benign & 9 & 1 & 11.1\% \\
Marketplace benign & 10 & 0 & 0.0\% \\
Support benign & 9 & 0 & 0.0\% \\
Benign finance/crypto & 9 & 1 & 11.1\% \\
Gift-card benign & 10 & 1 & 10.0\% \\
\midrule
\textbf{Total} & \textbf{76} & \textbf{5} & \textbf{6.6\%} \\
\bottomrule
\end{tabular}
\end{table}

\begin{table}[H]
\centering\small
\caption{Error taxonomy for false positives in the scam-adjacent benign stress test.}
\label{tab:scam_adjacent_benign_error_taxonomy}
\begin{tabular}{lcc}
\toprule
\textbf{False-positive type} & \textbf{Count} & \textbf{Rows} \\
\midrule
Benign wrong-number continuation & 2 & sab\_005, sab\_006 \\
Legitimate family money help & 1 & sab\_036 \\
Benign crypto/finance discussion & 1 & sab\_054 \\
Verified gift-card coordination & 1 & sab\_063 \\
\midrule
\textbf{Total} & \textbf{5} & -- \\
\bottomrule
\end{tabular}
\end{table}

\subsection{Scalability of Summary-Based Memory}
\label{sec:scalability}

One natural baseline for conversation-level detection is to hand the model an entire transcript at each decision point (full history), or to classify the whole conversation in a single pass (a full-transcript classifier). On short conversations both would perform comparably, since the transcript still fits a manageable context. Neither scales. The context handed to the detector grows with the conversation, so the cumulative input cost of re-reading the full history every few turns grows quadratically in the number of turns, and eventually a single full-transcript prompt exceeds the model's context window. Summary-based memory avoids this by design: it supplies a bounded running summary plus a short recent-message window at each decision, so the per-decision context stays bounded no matter how long the conversation runs.

\begin{figure}[t]
	\centering
	\includegraphics[width=0.82\linewidth]{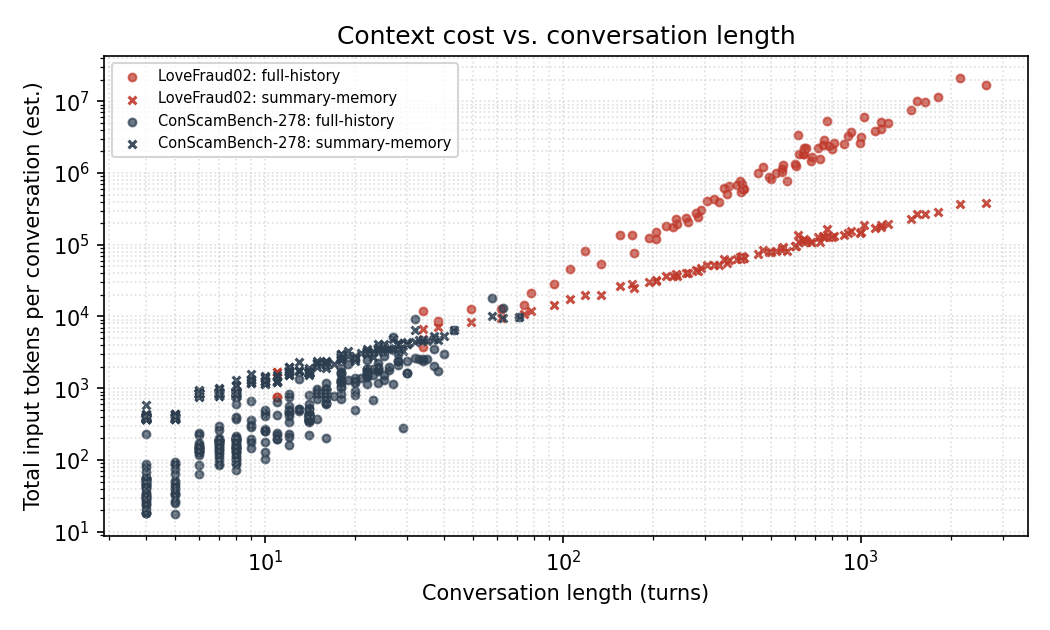}
	\caption{Per-conversation input-token cost (estimated) versus conversation length, for full-history prompting (circles) and the proposed summary-based memory (crosses), on both corpora. The two strategies coincide for short conversations but diverge sharply as conversations lengthen, with full-history growing super-linearly on the long LoveFraud02 dialogues while summary-based memory remains bounded. Both axes are logarithmic.}
	\label{fig:context-growth}
\end{figure}

We measured the input-context cost of both strategies on the two evaluation corpora, making one conversation-level decision every three turns. The measurement is a counting exercise and needs no model inference. We estimate token counts as characters/4, so absolute values are approximate while the ratios and growth trends hold regardless of tokenizer. Table~\ref{tab:scalability} reports the results and Figure~\ref{fig:context-growth} plots per-conversation input cost against conversation length.

\begin{table}[H]
	\centering\small
	\caption{Input-context cost of full-history prompting versus the proposed summary-based memory, as a function of conversation length. Tokens are estimated (chars/4); ratios are model-independent. A conversation-level decision is made every three turns.}
	\label{tab:scalability}
	{
		\setlength{\tabcolsep}{4pt}
		\begin{tabular}{lcc cc c cc}
			\toprule
			                 & \multicolumn{2}{c}{Turns} & \multicolumn{3}{c}{Avg. input tokens / conv.} & \multicolumn{2}{c}{Peak single prompt}                                                       \\
			\cmidrule(lr){2-3}\cmidrule(lr){4-6}\cmidrule(lr){7-8}
			Dataset          & Median                    & Max                                           & Full-history                           & Summary  & Reduction    & Full-transcript & Summary \\
			\midrule
			LoveFraud02      & 469                       & 2622                                          & 2{,}120{,}508                          & 93{,}707 & 22.6$\times$ & 58{,}138        & 2{,}547 \\
			ConScamBench-278 & 11 & 71 & 861 & 1{,}830 & 0.5$\times$ & 1{,}701 & 654 \\
			\bottomrule
		\end{tabular}
	}
\end{table}

On LoveFraud02, whose pre-scam conversations have a median of 469 turns and reach 2,622, full-history prompting consumes on average about $2.12$M input tokens per conversation against roughly $94$k for summary-based memory, a $22.6\times$ reduction, and the largest single full-transcript prompt reaches about $58$k tokens. Passed as a single transcript, $52$ of the $83$ conversations exceed an 8k-token context and $8$ exceed 32k, which makes the full-transcript classifier not only expensive but unusable for a substantial fraction of realistic long-form scams under common context limits. Summary-based memory keeps the peak prompt near $2.5$k tokens throughout. On the short conversations of ConScamBench-278 (median 11 turns) the picture reverses. Summary-based memory costs more on average than full history ($1{,}830$ versus $861$ tokens), since its fixed summary overhead never amortizes over so few turns. Section~\ref{sec:baselines} compares the full-transcript classifier's accuracy directly, where the proposed system is more accurate on both corpora. Here, we isolate the cost on which full-history prompting becomes prohibitive at scale. The opposite extreme, a fixed recent-window classifier, bounds the context by using only the final $k$-message window. It escapes the cost growth, but judges the conversation from a narrow local context and throws away earlier state. It cannot link an early cue to a much later one, an identity claim or promise introduced hundreds of turns before it is contradicted or exploited, where no single window is damning on its own. Summary-based memory occupies the point between the extremes: bounded in cost, unlike full history, and integrating evidence from the whole conversation into one evolving state, unlike a fixed window. These properties make it a suitable state representation for scams in long, evolving conversations, and since the investigator, the adaptive scheduler, and the advice generator all consume the same summary, for driving the explainable multi-agent pipeline as a whole.

\section{User Study}

We ran two user studies, A and B, to understand the problem space and to gather feedback on the proposed model. Both were conducted with informed consent. Together, they address the following research questions:

\begin{itemize}
	\item \textbf{RQ1. Need:} Do users' experiences with suspicious messages suggest a need for scam-detection support?
	\item \textbf{RQ2. Effectiveness Baseline:} How well can users identify scam-like conversations without AI assistance?
	\item \textbf{RQ3. Uncertainty:} Under what circumstances are users uncertain or divided when judging whether a conversation is a scam?
	\item \textbf{RQ4. Trust and Disagreement:} What is users' initial trust in AI-based scam-detection advice, how do their self-reported trust and confidence differ before versus after a guided session (interpreted as exploratory reception rather than a controlled effect), and under what circumstances do users agree or disagree with the proposed detector?
	\item \textbf{RQ5. AI Errors:} How do users respond when the detector provides an incorrect or imperfect scam-detection decision?
\end{itemize}

In Study~A, participants had no access to the system and were asked to detect scams in conversations unaided. The study examined their prior exposure to scam-like messages, their baseline ability to recognize scams, and their safety behavior in suspicious situations. Study~A addresses RQ1--RQ3 and part of RQ4. In Study~B, participants used the proposed system and evaluated several scam and non-scam conversations, giving feedback before and after use. We treat Study~B as exploratory usability and reception feedback, not a controlled evaluation. Study~B addresses RQ5 and part of RQ4. Table~\ref{tab:participant_demographics} summarizes the demographics and sample characteristics of both studies.

\subsection{Participant Demographics}

Study~A included 100 participants, Study~B included 45, and the two samples were disjoint. As Table~\ref{tab:participant_demographics} shows, most participants in both studies were between 25 and 44 (66\% in Study~A, 53\% in Study~B), with representation across the older age bands, including participants aged 55 and above. Most reported prior exposure to suspicious messages (67 of 100 in Study~A, 32 of 45 in Study~B), and daily text-messaging use was common. The spread across age and experience matters here, since susceptibility to conversational scams varies with age and digital familiarity.

\begin{table}[H]
	\centering
	\small
	\caption{Participant demographics and sample characteristics.}
	\label{tab:participant_demographics}
	\begin{tabular}{p{5.2cm}cc}
		\toprule
		\textbf{Characteristic}                 & \textbf{User Study A} & \textbf{User Study B} \\
		\midrule
		Number of participants                  & 100                   & 45                    \\
		\midrule
		Age 18--24                              & 8                     & 6                     \\
		Age 25--34                              & 50                    & 14                    \\
		Age 35--44                              & 16                    & 10                    \\
		Age 45--54                              & 14                    & 8                     \\
		Age 55--64                              & 12                    & 5                     \\
		Age 65 or older                         & 0                     & 2                     \\
		\midrule
		Uses text messaging daily               & 73                    & 30                    \\
		Previously received suspicious messages & 67                    & 32                    \\
		\bottomrule
	\end{tabular}
\end{table}

\subsection{RQ1. Need}

Table~\ref{tab:study-a-need-summary} presents results from Study~A on participants' prior experiences and self-assessments. Of the 100 consenting participants, 67\% had previously encountered suspicious or scam-like text messages and 17\% were unsure. Exposure to text-based communication was observed high where 73\% used SMS or text messaging at least daily. Mean self-rated scam-spotting confidence was 3.35 out of 5, and 43\% of participants rated their confidence at 3 or below. As such, a good many users feel uncertain when judging suspicious messages. Results show that scam messages are a common experience and these findings support the need for the proposed system.

\begin{table}[H]
	\centering
	\caption{Need-related findings from User Study A.}
	\label{tab:study-a-need-summary}
	\begin{tabular}{p{0.68\linewidth}c}
		\toprule
		\textbf{Metric}                               & \textbf{Value} \\
		\midrule
		Participants                                  & 100            \\
		Prior suspicious/scam-like exposure: Yes      & 67.0\%         \\
		Prior suspicious/scam-like exposure: Not sure & 17.0\%         \\
		SMS/text usage at least daily                 & 73.0\%         \\
		Mean scam-spotting confidence                 & 3.35/5         \\
		Confidence rated 3 or below                   & 43.0\%         \\
		Risky safety response                         & 17.0\%         \\
		\bottomrule
	\end{tabular}
\end{table}

Table~\ref{tab:safety-response-performance} shows how participants respond to risky scam-related situations, especially when conversations include pressure, secrecy, or urgent payment requests. 83.0\% of participants selected the safer response of pausing and verifying through an official or known channel, and 17.0\% selected risky actions, such as keeping the situation secret or sending payment quickly. This suggests a need to provide clear guidance on safe next steps.

\begin{table}[H]
	\centering
	\caption{Scam-detection performance by safety-response group in User Study A.}
	\label{tab:safety-response-performance}
	\begin{tabular}{lcccc}
		\toprule
		\textbf{Group} & \textbf{n} & \textbf{Strict Accuracy} & \textbf{Unsure} & \textbf{Decided Accuracy} \\
		\midrule
		Risky response & 17         & 62.25\%                  & 22.06\%         & 79.44\%                   \\
		Safer response & 83         & 61.35\%                  & 21.49\%         & 77.86\%                   \\
		\bottomrule
	\end{tabular}
\end{table}

\subsection{RQ2. Effectiveness Baseline}

Table~\ref{tab:baseline-performance} shows participants' scam-detection performance across 12 conversations. With 100 participants, we collected 1,200 scam and non-scam judgments. Treating ``Not Sure'' responses as incorrect, accuracy was observed as 61.5\% (95\% CI [58.7, 64.2]). Counting only definite Yes/No decisions raises accuracy to 78.4\% (95\% CI [75.7, 80.9]), but 21.6\% of all responses were ``Not Sure,'' which points to substantial uncertainty. Participants identified scam conversations correctly in 73.5\% of cases and recognized non-scam conversations as safe in only 49.5\%. They often reached the right conclusion while uncertainty and false positives stayed common, which again argues for detection tools of the kind proposed here.

\begin{table}[H]
	\centering
	\caption{Baseline scam-detection performance in User Study A. Proportions are shown with 95\% Wilson confidence intervals.}
	\label{tab:baseline-performance}
	\begin{tabular}{lc}
		\toprule
		\textbf{Metric}                             & \textbf{Value (95\% CI)} \\
		\midrule
		Total judgments                             & 1,200                    \\
		Strict accuracy                             & 61.5\% [58.7, 64.2]      \\
		Decided accuracy                            & 78.4\% [75.7, 80.9]      \\
		``Not sure'' responses                      & 21.6\% [19.4, 24.1]      \\
		Scam conversations correctly identified     & 73.5\%                   \\
		Non-scam conversations correctly identified & 49.5\%                   \\
		\bottomrule
	\end{tabular}
\end{table}

\subsection{RQ3. Uncertainty}

\begin{figure}[H]
	\centering
	\includegraphics[width=\linewidth]{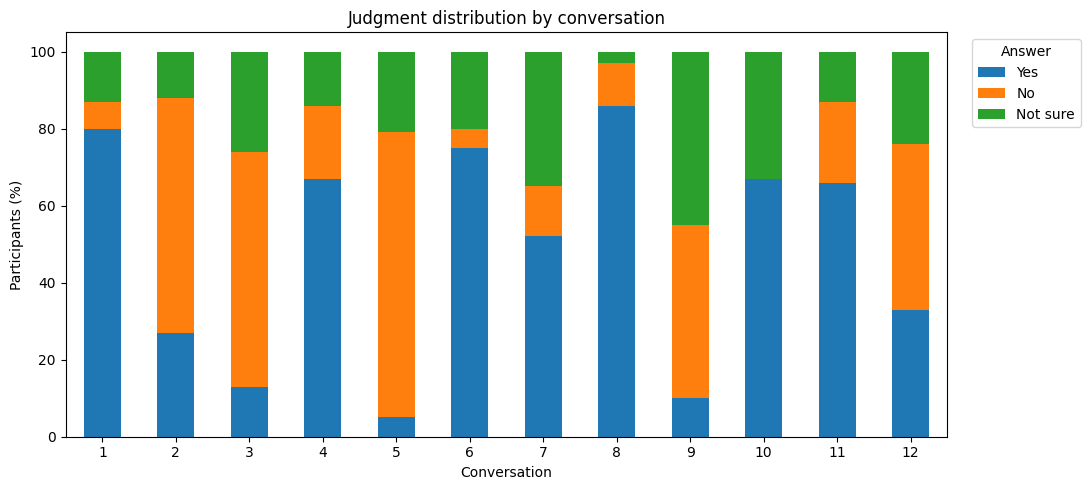}
	\caption{Distribution of participant judgments across the 12 conversations in User Study A, highlighting the percentage of participants who answered ``Yes,'' ``No,'' or ``Not sure'' when judging whether the conversation was a scam.}
	\label{fig:judgment-distribution}
\end{figure}

Figure~\ref{fig:judgment-distribution} shows the distribution of participant judgments across the 12 conversations. Uncertainty (green) concentrates in a few conversations rather than spreading evenly across scenarios. By both the proportion of ``Not Sure'' responses and answer entropy, the most uncertain and divisive cases were often the non-scam conversations. Answer entropy is computed per conversation over the empirical distribution of responses across the three options \{Yes, No, Not sure\}. Letting $p_i$ be the proportion of responses assigned to option $i$, the normalized answer entropy is $H_{\mathrm{norm}} = -\frac{1}{\log 3}\sum_{i=1}^{3} p_i \log p_i$ (omitting $p_i = 0$ terms), which maps complete agreement to 0 and maximal disagreement across the three options to 1. Participants struggled both to identify scams and to see when suspicious-looking interactions were in fact benign. Users appear to grow overly cautious in ambiguous situations, and the proposed system can help them separate genuine threats from harmless conversations.

\subsection{RQ4. Trust and Disagreement}

We compared participants' self-reported ratings before and after they used the system. Before use, initial trust in AI safety advice was mixed (mean 3.89 out of 5). In this uncontrolled, single-session pre/post design, ratings rose across every measured dimension. Each increase was statistically significant by a one-sided Wilcoxon signed-rank test. Trust from 3.89 to 4.60 ($p<0.001$), self-confidence from 3.42 to 4.33 ($p<0.001$), belief that AI can help detect scams from 3.91 to 4.67 ($p<0.001$), and perceived need for an AI scam detector from 3.42 to 4.24 ($p<0.001$). On every dimension, all non-tied participant-level changes were increased (rank-biserial correlation $r=1.00$) and no participant's rating fell. We read these pre/post ratings as exploratory reception data rather than a controlled measurement of trust formation, since the design has no control condition and collects ratings immediately after a single guided session. The uniformity itself is suspicious. Not one participant's rating decreased on any of the four measures ($r=1.00$ throughout), which fits demand, acquiescence, or ceiling effects better than a real treatment effect. We therefore report the numbers only as evidence that the users received the system positively. Also, we lean on the behavioral measures below (agreement and post-warning actions) and on User Study~A for more defensible user-facing evidence.

Table~\ref{tab:study-b-agreement-summary} reports participant agreement with the system. Across 540 conversation-level judgments, participants at least partially agreed with the detector in 86.3\% of cases (95\% CI [83.1, 88.9]). Agreement was similar whether the system said ``Scam'' (87.8\%, 95\% CI [83.3, 91.2]) or ``Not Scam'' (84.8\%, 95\% CI [80.0, 88.6]), so it did not simply reflect users deferring to alarming warnings.

\begin{table}[H]
	\centering
	\caption{Participant agreement with the proposed system across User Study B conversation judgments.}
	\label{tab:study-b-agreement-summary}
	\begin{tabular}{lccc}
		\toprule
		\textbf{System Output} & \textbf{Yes} & \textbf{Partially} & \textbf{No} \\
		\midrule
		Scam                   & 66.3\%       & 21.5\%             & 12.2\%      \\
		Not Scam               & 64.8\%       & 20.0\%             & 15.2\%      \\
		Overall                & 65.6\%       & 20.7\%             & 13.7\%      \\
		\bottomrule
	\end{tabular}
\end{table}

\subsection{RQ5. AI Errors}

Participants did not treat the system as an unquestionable authority. Of 270 scam warnings, about 33.7\% of judgments involved partial agreement or disagreement. Contested judgments did not always lead to unsafe behavior. When the system produced a scam warning, 86.7\% of responses still chose a safe action (95\% CI [82.1, 90.2]), as Table~\ref{tab:behavior_after_warnings} shows. Among partial agreements with scam warnings, 89.7\% chose a safe action and no participant continued the conversation after the warning. The residual-risk group is the participants who explicitly disagreed with scam warnings. 66.7\% chose a safe action and 21.2\% kept the conversation going until the warnings grew more obvious. Most responses, then, were consistent with safer decision-making. Informal feedback asked for stronger escalation cues and clearer next-step guidance when users reject high-risk scam warnings.

\begin{table}[H]
	\centering
	\caption{Participant behavior after scam-detector warnings in User Study B, with 95\% Wilson confidence intervals on the safe-action rate.}
	\label{tab:behavior_after_warnings}
	\begin{tabular}{lrccc}
		\toprule
		Group                          & n   & Safe Action (95\% CI) & Continued & Not Sure \\
		\midrule
		All scam warnings              & 270 & 86.7\% [82.1, 90.2]   & 7.4\%     & 5.9\%    \\
		Agreement with warning         & 179 & 89.4\% [84.0, 93.1]   & 7.3\%     & 3.4\%    \\
		Partial agreement with warning & 58  & 89.7\% [79.2, 95.2]   & 0.0\%     & 10.3\%   \\
		Disagreement with warning      & 33  & 66.7\% [49.6, 80.2]   & 21.2\%    & 12.1\%   \\
		\bottomrule
	\end{tabular}
\end{table}

\begin{figure}[t]
	\centering
	\includegraphics[width=0.8\linewidth]{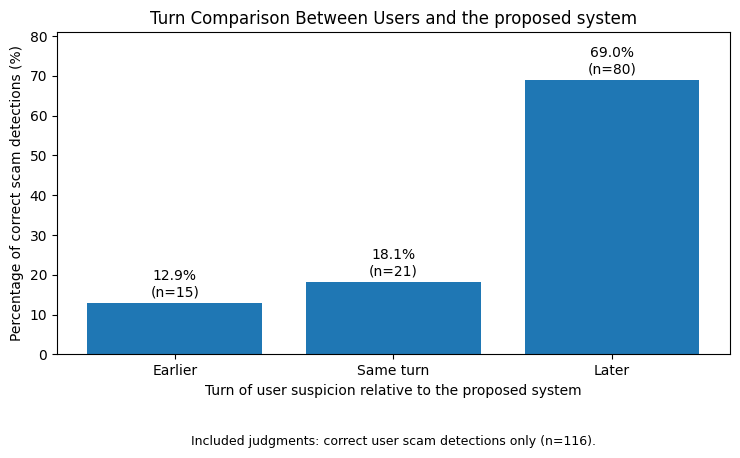}
	\caption{Whether participants identified the scam earlier than the detector, on the same turn, or later.}
	\label{fig:timing-comparison}
\end{figure}

Figure~\ref{fig:timing-comparison} covers the cases where participants ultimately judged correctly, answering ``Yes'' to whether the conversation was a scam. 69.0\% of correct detections came later than the system's detection turn, 18.1\% on the same turn, and 12.9\% earlier. Most participants needed more conversational evidence than the system before identifying the scam. The remaining 31.0\% of correct judgments, on the same turn as the system or earlier, came from participants who took a more cautious approach.

\subsection{Perceived Usability}
\label{sec:usability}

To assess practical usability, Study~B participants completed the ten-item System Usability Scale (SUS)~\cite{brooke1996sus} after their session where an eleventh attention-check item served only for data-quality screening. The system scored a mean of 74.7 (SD 7.6, 95\% CI [72.5, 76.9]), above the widely used benchmark of 68 that separates below from above-average usability and corresponds to a ``good'' adjective rating on the interpretive scale of Bangor et al.~\cite{bangor2008sus}. 80\% of participants individually rated the system at or above 68. The explainability and advice-generation layers thus produce a system non-expert users find usable, which complements the detection-accuracy results above.

\section{Limitations and Future Work}
\label{sec:limitations}

The proposed system has several limitations. First, it still produces occasional false positives, particularly on benign wrong-number conversations that resemble scam grooming. On the expanded ConScamBench-278 benign set (150 non-scam conversations) the false-positive rate is a low 2.7\% (95\% CI [1.0, 6.7]), but lowering it further without sacrificing recall would be worthwhile. The system can also miss borderline scams that never escalate into a concrete malicious action: the two missed ConScamBench-278 scams were real conversations that ended before any payment, artifact, or off-platform redirection materialized. More fundamentally, our estimate of false-positive behavior on scam-adjacent benign conversations comes from an authored stress test (Section~\ref{sec:stress_test}: 76 benign conversations that imitate scam entry channels, false-positive rate 6.6\%, 95\% CI [2.8, 14.5]), not from a naturally occurring corpus, and LoveFraud02, our one naturally occurring corpus, is all-scam. The stress test sits close to the decision boundary by design, and its dominant failure mode is again benign wrong-number continuations. Establishing deployable false-alarm rates on live traffic will require a real benign conversational corpus drawn from scam-bearing channels, which we leave to future work.

Second, the evaluation datasets may not capture the full diversity of real-world scams. LoveFraud02 mainly represents romance fraud, and ConScamBench-278, broad as its category coverage is, remains a controlled benchmark rather than a sample of naturally occurring benign conversations. Because 18 of its 128 scam conversations were generated with an LLM, and because the SMD comparison uses an LLM backend, optimistic bias on that synthetic portion is possible. We mitigate it by checking that detection also holds on the real-sourced majority and by evaluating on the naturally occurring LoveFraud02 corpus, though a larger, fully human-sourced conversational benchmark would still be valuable. The detection architecture is the central contribution; ConScamBench-278, released publicly across eight scam families, is meant to be extended and refined as the community assembles larger conversational corpora.

Third, no prior open detector exists for multi-turn conversational scam detection, so we anchor the CSD against a single-message baseline, unassisted human performance (User Study~A), component attribution (Table~\ref{tab:detection_source_distribution}), a component ablation (Section~\ref{sec:ablation}), and a recent-window conversational baseline (Section~\ref{sec:baselines}), and we check its behavior across two open-weight backends (Table~\ref{tab:backend_portability}). The recent-window comparison gives an empirical bounded-context reference for the proposed summary-based memory; the full-history analysis in Section~\ref{sec:scalability} explains why re-supplying the entire transcript does not scale. Future work can use ConScamBench-278 for head-to-head comparisons with later conversational scam detectors and to test other window sizes or memory strategies.

Fourth, we evaluated the system in a controlled testbed rather than a live messaging environment, so the conversations do not reflect the interruptions, latency, or missing context of deployment, and the user-study gains in trust were measured immediately after a single session rather than longitudinally.

Fifth, summary-based memory bounds context cost (Section~\ref{sec:scalability}) but may omit subtle details that accumulate over a long conversation, and the adaptive schedule may delay warnings in fast-moving scams. The token-cost analysis uses an estimated tokenizer (chars/4), so absolute costs are approximate, though the scaling trends do not depend on the tokenizer. The privacy-oriented design keeps all conversation content and user data local, yet the system is not fully self-contained: brand-domain verification issues lookups to Serper.dev, a third-party search service, transmitting isolated brand names and artifacts and no conversation context. Removing that residual external call would require a self-hosted search index or a cached brand-to-domain allowlist. Our user-facing advice is evidence-based, but we have not formally evaluated its explanation faithfulness, whether the stated rationale reflects the cues the detector actually used; that too is future work. Finally, the multilingual advice capability needs deeper evaluation for cross-language consistency.

\section{Conclusion}
Detection tools so far have focused on isolated messages rather than multi-turn manipulative conversations. This paper proposes an explainable agentic system for conversational scam detection. It combines single-message detection with conversation-level detection to catch both explicit malicious artifacts and scams that emerge across a series of dialogue turns. We evaluate it on the public LoveFraud02 dataset and on our own curated multi-category dataset, ConScamBench-278. The system reaches 100\% conversational-scam recall on LoveFraud02 corpus, which we treat as our primary evidence of real-world detection recall (being an all-scam corpus, LoveFraud02 says nothing about false-alarm behavior), and 97.8\% accuracy (95\% CI [95.4, 99.0]) on ConScamBench-278, an initial public multi-category conversational-scam benchmark. This benchmark mostly composes real-sourced with minority of adversarial LLM-generated simulations, making it a strong controlled complement to the LoveFraud02 evidence across eight scam families. We will release ConScamBench-278 for further research, along with the agent prompts, the orchestrator logic, the evaluation harness needed to reproduce our results, and the per-message label diffs for the SMD dataset. The recent-window comparison shows that summary-based memory preserves early conversational evidence that a final bounded window would discard. The proposed system matches the full-transcript classifier's accuracy on ConScamBench-278, exceeds it on LoveFraud02, catches more scams on both corpora, and does so at a marginally higher false-positive rate for a fraction of the cost. The system distills the dialogue into a compact evolving memory which is as effective as supplying the raw transcript, and far cheaper. A component ablation confirms that removing the summary memory lowers recall and delays detection. The behavior also transfers across two open-weight backends, so it is a property of the architecture rather than of one model, which supports the privacy-oriented, self-hosted design. Two user studies are carried out to round out the evaluation. Participants often struggle with uncertainty when judging suspicious conversations. After a controlled, single-session interaction with the system they reported higher trust, self-confidence, and perceived need for AI scam detection, and rated usability above the standard SUS benchmark. Responses were collected without a control condition and immediately after use. As such, the measures indicate user reception rather than trust formation. The results point to the value of AI scam-detection support in helping users reconsider judgments they are unsure about.

\section{Bibliography}
\section*{Appendix}
\section*{A. Reproducibility Details}
\label{sec:reproducibility}

\paragraph{Models and inference.} Every text agent runs on deepseek-v3.1:671b, served through an OpenAI-compatible endpoint (Ollama Cloud, \texttt{https://ollama.com/v1}). The backend-portability check re-runs the conversation-level detector on Qwen2.5-72B-Instruct. When enabled, the single-message screenshot step uses the qwen3-vl:30b vision model on a local Ollama instance, and the pipeline degrades gracefully if it is unavailable. Calls use temperature $0$ and structured JSON output (\texttt{response\_format: json\_object}, with an automatic fallback for hosts that reject it). No other decoding parameters were modified. At temperature $0$ inference is deterministic, and a disk cache keyed on the prompt makes runs reproducible and resumable.

\paragraph{Orchestration and adaptive schedule.} The artifact trigger runs only on messages from the external party (the ``Sender''), never on user-authored messages. The investigator and the conversation-level detector first runs once a conversation reaches three messages. This is the minimum length we treat as a conversation. After that, the cadence follows the threat level held in the running summary. The detector re-runs on every turn at high, on every second turn at medium, on every third at low, and on every fifth at none.

\paragraph{Decision rule and latency.} A conversation counts as a scam if either the single-message detector or the conversation-level detector fires at any point. The first-detection turn is the earliest turn at which either path fires, and reported detection-latency averages cover detected conversations only.

\paragraph{Verification.} Before final classification, the verifier extracts URLs, phone numbers, email addresses, and claimed brand identities, and checks them. A local headless browser retrieves URL content, which Docling then parses, in place of the third-party Jina Reader used by the baseline. The brand-to-official-domain lookup goes through the Serper.dev search API (Section~\ref{stage1}), which receives isolated brand names and artifacts only.

\paragraph{Release.} We release the structured-output schemas, orchestration and verification logic, the evaluation harness (single-message, conversational, ablation, backend-portability, and scalability scripts), ConScamBench-278, and the per-message SMD label diffs, so that every reported result can be reproduced.

\bibliographystyle{unsrt}
\bibliography{refs}
\end{document}